\documentclass[aps,pra,showpacs,floatfix,amsmath,amssymb,longbibliography,10pt]{revtex4-1}

\usepackage{graphicx}
\usepackage{dcolumn}
\usepackage{bm}
\usepackage{amssymb}
\usepackage{amsmath}
\usepackage{float}
\usepackage{multirow}
\usepackage{tgtermes}
\usepackage{ulem}
\usepackage{color}
\usepackage[english]{babel}
\usepackage{txfonts}
\usepackage[altg]{qtxmath}
\usepackage{microtype}
\usepackage{slashed}
\usepackage{diagbox}

\renewcommand*{\vec}[1]{\boldsymbol{#1}}

\makeatletter
\newcommand*{\rom}[1]{\expandafter\@slowromancap\romannumeral #1@}
\makeatother

\begin{document}

\author{Q.~Z. Lv}
\email{qingzheng.lyu@mpi-hd.mpg.de}
\author{E.~Raicher}
\thanks{Present Address: Soreq Nuclear Research Center, Yavne 80800, Israel}
\email{erez.raicher@mail.huji.ac.il}
\author{C.~H. Keitel}
\author{K.~Z. Hatsagortsyan}
\affiliation{Max-Planck-Institut f\"{u}r Kernphysik, Saupfercheckweg 1,  69117 Heidelberg, Germany }



\title{Ultrarelativistic electrons in counterpropagating laser beams}

\begin{abstract}

The dynamics and radiation of ultrarelativistic electrons in strong counterpropagating laser beams are investigated.  Assuming that the particle energy is the dominant scale in the problem, an approximate solution of classical equations of motion is derived and the characteristic features of the motion are examined.  A specific regime is found with comparable strong field quantum parameters of the beams, when the electron trajectory exhibits ultrashort spike-like features, which bears great significance to the corresponding radiation properties.  An analytical expression for the spectral distribution of spontaneous radiation is derived in the framework of the Baier-Katkov semiclassical approximation based on the classical trajectory.  All the analytical results are further validated by exact numerical calculations.  We consider a non-resonant regime of interaction, when the laser frequencies in the electron rest frame are far from each other, avoiding stimulated emission.  Special attention is devoted to settings when the description of radiation via the local constant field approximation fails and to corresponding  spectral features.  Periodic and  non-periodic regimes are considered, when lab frequencies of the laser waves are always commensurate. The sensitivity of spectra with respect to the electron beam spread, focusing and finite duration of the laser beams is explored.

\end{abstract}

\maketitle

\section{Introduction}
\label{sec:intro}

Electromagnetic processes in strong laser fields are characterized by nonperturbative multiphoton dynamics. An efficient treatment of nonlinear processes in strong field quantum electrodynamics (QED) has been provided within the Furry picture \cite{Furry_1951}, regarding the strong field as classical and employing the electron wave function in such fields for the calculation of amplitudes of QED processes. The  Volkov wave function  of an electron in a plane wave laser field \cite{Volkov_1935} has been successfully and extensively employed to explore
the nonlinear Compton effect, nonlinear Breit-Wheeler \cite{Nikishov_1964,Ritus_1964,Ritus_1985}, and nonlinear Bethe-Heitler pair production processes \cite{Yakovlev_1966}. The multiphoton processes in a plane wave field enter into play at large values of the classical strong field parameter $\xi \equiv -ea/m\gg 1$, where $a$ is the amplitude of the vector potential, $a\equiv \sqrt{ A^2}$, while $e$ and $m$ the electron charge and mass, respectively. Relativistic units $\hbar=c=1$ are used throughout the paper, unless specified otherwise. Present day laser facilities attain intensities of up to $5 \times 10^{22} W/cm^2$ in optical wavelengths \cite{Yoon_2019, Vulcan1}, corresponding to $\xi \sim 100$. For the next generation extreme laser infrastructures an order of magnitude increase of intensity is expected \cite{ELI,XCELS}, opening a bright avenue for investigation of extreme nonlinear strong field QED processes  \cite{Marklund_2006,Mourou_2006,Dunne_2009,RMP_2012} in laser-plasma or laser-electron beam interactions.

The desire to increase the effective laser field with a given laser beam energy
gave rise to the concept of multi-beam configurations and to the notion of a dipole wave \cite{Bulanov_2010_a,Golla_2012,Gonoskov_2012,Gonoskov_2017,Bashinov_2013,Bashinov_2019,Magnusson_2019}. The simplest case of a multi-beam configuration is the counterpropagating laser beam setup, which is an attractive setup to study QED effects \cite{Kirk_2009,Bulanov_2010,Gonoskov_2014,Gong_2017,Grismayer_2017,Grismayer_2016,Kirk_2016,Jirka_2016,milosevic_2004use,hatsagortsyan_2006microscopic}.
All of the above admit no exact analytical solutions for the wave function and are, therefore, not accessible to strong field QED calculations within the Furry picture. The common way of treating strong field QED processes in laser-plasma interaction is to approximate the emission by that in the presence of the local constant field when the field intensity is very high ($\xi\gg 1$). The local constant field approximation (LCFA) is rigorously derived in the asymptotic limit $\xi\gg 1$ for the plane wave case (more precise condition is $ (\xi/\chi^{1/3}) [\omega/(\varepsilon - \omega)]^{1/3}\gg 1$ \cite{DiPiazza_2018,Blackburn_2020}, with typical emission frequencies $\omega/\varepsilon\sim \chi/(\chi+1)$ and the electron energy $\varepsilon$). In this case
the formation length of the process becomes smaller than the field wavelength, and the process probability depends solely on the quantum parameter $\chi= e\sqrt{-(F^{\mu \nu} P_{\nu})^2}/m^3$, where $F_{\mu \nu} $ is the electromagnetic tensor and $P_{\mu}=(\varepsilon,\textbf{P})$ the particle 4-momentum. Due to its simplicity, this approximation allows for the inclusion of QED processes in kinetic Monte Carlo and particle-in cell (PIC) simulations involving fields of complex forms  \cite{Elkina_2011,Ridgers_2014,Green_2015}.  However, recently deficiencies and failures of LCFA have been observed in low \cite{DiPiazza_2018,DiPiazza_2019,Ilderton_2019a,Ilderton_2019b} and high energy limits \cite{Podszus_2019}.  LCFA violation in counterpropagating laser waves is demonstrated in \cite{lv2021_anomalous} which is due to emergence of an additional small time scale in the electron dynamics.

Beyond LCFA treatment, one may apply the Wentzel-Kramers-Brillouin (WKB) approximation to describe the electron quantum (quasiclassical) dynamics \cite{Popov_1997,Mocken_2010}.
A similar high-energy approximation describing the electron dynamics in a focused laser field, when the electron longitudinal momentum dominates over transverse one, is developed in \cite{DiPiazza_2014} and applied for description of corresponding nonlinear QED processes \cite{DiPiazza_2015,DiPiazza_2016,DiPiazza_2017}. As WKB approximation is closely connected with the classical description, a WKB wave function in closed analytical form can be derived in the cases when such solution is available for the electron classical trajectory. In the 60s' this observation motivated Baier and Katkov to develop the operator approach and with its help to express the amplitudes of strong field QED processes, such as radiation and pair production,  as a function of the electron classical trajectory in the external field \cite{Katkov_1968,Baier_b_1994, Landau_4}.


We consider the setup of counterpropagating laser beams. Here one should distinguish resonant and non-resonant regimes of interaction. The resonance appears when the frequencies of the laser waves match in the average rest frame of the electron \cite{Avetissian_b_2016}, which would lead to stimulated emission of laser photons \cite{Friedman_1988,Fedorov_1981},
to coherent electron scattering from the moving laser grating (Kapitza-Dirac effect \cite{Kapitza_1933,Batelaan_2007,Ahrens_2012,Mueller_2017}).
Rather than the widely explored topic of stimulated processes in the resonant regime, we discuss in this paper the non-resonant regime, relevant to the investigation of spontaneous radiation in this setup. The equation of motion is highly nonlinear and is known to exhibit chaotic dynamics when the corresponding field are strong \cite{Lehmann_2012,Bashinov_2015}. In the quantum domain, approximations to the wave function of a scalar particle experiencing this field have been discussed in \cite{Hu_2015,King_2016}. Radiation in this setup and its reaction to the electron dynamics have been  investigated within LCFA via PIC-QED simulations \cite{Kirk_2009,Grismayer_2016,Jirka_2016,Gong_2017,Grismayer_2017}. In particular, this configuration turned out to be favorable to QED cascades where the emitted $\gamma$-photons are energetic enough to produce electron-positron pairs, starting an avalanche-like dynamics. Moreover, it was shown that radiation reaction can essentially modify the trapping of particles in this field \cite{Gonoskov_2014,Kirk_2016}.

In the present paper an electron interacting with counterpropagating laser beams in the non-resonant regime is considered, 
using laser fields of equal frequency in the laboratory frame and the ultrarelativistic electron moving initially along the propagation direction of the first laser beam. An approximate analytical solution to the classical equation of motion is derived, imposing a restriction on the laser parameters and electron initial momentum, in particular, demanding $\xi_1 \xi_2\ll \gamma^2$, for the lasers' field parameters $\xi_1, \xi_2$ and $\gamma$ as the average Lorentz factor of the electron in the fields.  Based on the approximated analytical trajectory, the radiation is calculated in the realm of the semiclassical Baier-Katkov formalism.  We compare the obtained formula with a fully numerical calculation and discuss radiation features in different regimes. Furthermore, the influence of the pulses width and focusing, which cannot be accounted for analytically, are studied numerically.

The paper is organized as follows. In Sec.~\ref{sec:clasdyn} an approximate solution to the Lorentz equation in the counterpropagating beams is derived.
The investigation of the photon emission is given in Sec.~\ref{sec:radcal}.
Radiation spectra in strong fields are discussed along with a numerical example.
The validity of the analytical treatment and the deviations with respect to numerical calculations are analyzed.
The impact of finite duration and focusing of the laser beam is investigated numerically. Conclusions are given in Sec.~\ref{sec:con}.

\section{The classical dynamics}
\label{sec:clasdyn}

The classical equation of motion for the particle in  electromagnetic (EM) fields reads
\begin{equation}
\frac{d \textbf{P}}{d \tau} = e \gamma \left[ \textbf{E} \left( \tau, \textbf{x}(\tau)  \right) + \textbf{v} \times \textbf{B} \left( \tau, \textbf{x}(\tau)  \right)  \right],
\label{eq:av EOM}
\end{equation}
where $\tau$ is the proper time, $\gamma=\varepsilon/m$ is the relativistic Lorentz-factor, $\varepsilon = \sqrt{m^2+\textbf{P}^2}$ is the energy, $\textbf{P}$ is the momentum, $\textbf{E},\textbf{B}$ are the electric and magnetic fields, correspondingly, and $\textbf{v} = \textbf{P} / \varepsilon$ is the velocity.  In the general case, Eq.~(\ref{eq:av EOM}) cannot be solved analytically because of its nonlinearity, as
$\textbf{x}(\tau)$ depends on the momentum via $\textbf{x}(\tau)= \int d \tau  \textbf{P}(\tau)/m$.
In the following, we seek for an approximated solution in the presence of counterpropagating circularly polarized laser waves with the four-vector potential $A=A_1+A_2$ where
\begin{equation}
 \label{eq:av A_def}
A^{\mu}_1 \equiv a_1 g_1(k_1 \cdot x) \left[ \cos (k_1 \cdot x) e^{\mu}_x +
\sin (k_1 \cdot x) e^{\mu}_y
 \right] \, ; \,
A^{\mu}_2  \equiv  a_2 g_2(k_2 \cdot x) \left[ \cos (k_2 \cdot x) e^{\mu}_x +
\sin (k_2 \cdot x) e^{\mu}_y
 \right].
\end{equation}
The four-wavevectors of the beams are $k_1 = (\omega,0,0,\omega), k_2 = (\omega,0,0,-\omega)$ and $e_x = (0,1,0,0), e_y = (0,0,1,0)$ are the unit vectors. The dimensionless functions  $g_1(k_1 \cdot x)$ and $g_2(k_2 \cdot x)$ are slow wave envelopes. In this section they will be set to unity. We will refer to them when considering the influence of the turn-on process on the relation between the average momentum and its initial value in Sec.~\ref{sec:PzIN}.   Here $a \cdot b$ denotes the inner product of two four-vectors. The electric and magnetic fields are derived from the vector potential through $\textbf{E} = -\frac{\partial \textbf{A}}{\partial t}$ and
$\textbf{B} = \nabla \times \textbf{A}$:
\begin{equation}
\label{eq:efield}
\textbf{E}_1 = -a_1 \omega \left[ -\sin \left( k_1 \cdot x \right) \textbf{e}_x + \cos \left( k_1 \cdot x \right) \textbf{e}_y \right]\,; \,
\textbf{E}_2 = -a_2 \omega \left[ -\sin \left( k_2 \cdot x \right) \textbf{e}_x + \cos \left( k_2 \cdot x \right) \textbf{e}_y \right] \,.
\end{equation}

\begin{equation}
\label{eq:bfield}
\textbf{B}_1 = a_1 \omega \left[ \cos \left( k_1 \cdot x \right) \textbf{e}_x + \sin \left( k_1 \cdot x \right) \textbf{e}_y \right] \,; \,
\textbf{B}_2 =-a_2 \omega \left[ \cos \left( k_2 \cdot x \right) \textbf{e}_x + \sin \left( k_2 \cdot x \right) \textbf{e}_y \right] \,.
\end{equation}
Please note that here we have chosen the counterpropagating waves being co-rotating. However, the characterization of the electron dynamics is similar for counter-rotating waves in the considered regimes. The only difference is that the rotation caused by the $\xi_2$-beam changes its directions but the physical properties of radiation remain the same.

\subsection{Classical trajectory}
\label{sec:clastra}

For solving the equation of motion Eq.~(\ref{eq:av EOM}) the phases appearing in the fields arguments are expressed via the trajectory$\textbf{x}(\tau)$
\begin{eqnarray}
\phi_1 (\tau) \equiv  k_1 \cdot x(\tau) =   \frac{k_1 \cdot \bar{P}}{m} \tau + \delta \phi_1 (\tau), \quad
\phi_2 (\tau) \equiv  k_2 \cdot x(\tau) =   \frac{k_2 \cdot \bar{P}}{m} \tau + \delta \phi_2 (\tau)
\label{eq:av phi2Def}
\end{eqnarray}
where
\begin{equation}
\delta \phi_1 \equiv  \int{d\tau} \frac{k_1 \cdot \delta P (\tau)}{m},
\quad
\delta \phi_2 \equiv  \int{d\tau} \frac{k_2 \cdot \delta P (\tau)}{m},
\label{eq:av delPhi}
\end{equation}
with $\delta P_{\mu}=P_{\mu}(\tau)-\bar{P}_{\mu}$. The bar symbol designates time-averaged quantities.
The key assumption lying in the basis of our derivation is
\begin{equation}
\label{eq:key}
\int \sin \phi_1 d \tau \approx-\frac{m}{k_1 \cdot \bar{P}} \cos \phi_1 \, , \, \int \sin \phi_2 d \tau \approx -\frac{m}{k_2 \cdot \bar{P}} \cos \phi_2 \, , \, \int \sin (\phi_1-\phi_2) d \tau \approx - \frac{m}{(k_1-k_2) \cdot \bar{P}} \cos (\phi_1-\phi_2) ,
\end{equation}
as well as similar relations where in the right wing $\cos \rightarrow \sin$ and in the left wing $\sin \rightarrow -\cos$.
By employing this assumption, the 4-momentum $P$ of the particle can be derived.
With the momentum, expressions for $\delta \phi_1, \delta \phi_2$ according to Eq.~(\ref{eq:av delPhi}) are
calculated under certain restrictions, which  assure the validity of the assumption of Eq.~(\ref{eq:key}).

Since the vector potential is independent on the transverse coordinates, the canonical momentum in these directions is conserved $P_{\bot}(\tau) = p_{\bot}-eA(\tau)$.
Without loss of generality, we choose the initial transverse momentum $p_{\bot}$ to be on the $x$-axis. Then,
\begin{eqnarray}
P_x(\tau)&=& p_x + m \xi_1 \cos \phi_1 +m \xi_2 \cos \phi_2 \,,\label{eq:av final_Px}\\
P_y(\tau)&=&  m \xi_1 \sin \phi_1 +m \xi_2 \sin \phi_2 \,,
\label{eq:av final_Py}
\end{eqnarray}
where $-ea_{1,2}=m\xi_{1,2}$.
Applying the assumption of Eq.~(\ref{eq:key}), the $x,y$ components of the trajectory read
\begin{eqnarray}
x(\tau) &=&  \left(\frac{p_x}{m} \tau+ \frac{m \xi_1}{k_1 \cdot \bar{P}} \sin \phi_1 + \frac{m \xi_2}{k_2 \cdot \bar{P}} \sin \phi_2 \right) \,,
\label{eq:av final_x}\\
y(\tau) &=& - \left( \frac{m \xi_1}{k_1 \cdot \bar{P}} \cos \phi_1 + \frac{m \xi_2}{k_2 \cdot \bar{P}} \cos \phi_2 \right) \,.
\label{eq:av final_y}
\end{eqnarray}
Now let us consider the oscillations on the $z$ axis
\begin{equation}
 \frac{d \textbf{P}_z}{d \tau} = \frac{e}{m} \textbf{P}_{\bot} \times \textbf{B}.
\end{equation}
Employing Eqs.~(\ref{eq:av final_Px}), (\ref{eq:av final_Py}) and \eqref{eq:bfield},
one can find out that the terms scaling like $\xi_1^2, \xi_2^2$ cancel. Therefore we have
\begin{equation}
\label{eq:av dPz}
 \frac{d \textbf{P}_z}{d \tau} =
-p_x \omega \left[ \xi_1 \sin \phi_1 - \xi_2 \sin \phi_2 \right]
  -2 \omega m \xi_1 \xi_2  \sin \left( \phi_1-\phi_2 \right).
\end{equation}
Accordingly, $P_z = \bar{P}_{z}+ \delta P_z$ with
\begin{equation}
\delta P_z   =
p_x \omega \left[ \frac{m \xi_1}{k_1 \cdot \bar{P}} \cos \phi_1 - \frac{m \xi_2}{k_2 \cdot \bar{P}} \cos \phi_2 \right] +\frac{ 2m^2 \xi_1 \xi_2 \omega}{ (k_1-k_2) \cdot \bar{P}} \cos \left( \phi_1-\phi_2 \right),
\label{eq:av final_Pz}
\end{equation}
where $\bar{P}$ is the time-averaged momentum, whose relation to the initial momentum of the electron before interacting with the laser pulses will be discussed in Sec.~\ref{sec:PzIN}.
Integrating over $\tau$, one  obtains the $z$-component of the trajectory
\begin{equation}
z(\tau) = \frac{\bar{P}_z}{m} \tau
+\frac{ 2m^2 \xi_1 \xi_2 \omega }{ [(k_1-k_2) \cdot \bar{P}]^2} \sin \left( \phi_1-\phi_2 \right)
+ p_x \omega \left[ \frac{m \xi_1}{(k_1 \cdot \bar{P})^2} \sin \phi_1 - \frac{m \xi_2}{(k_2 \cdot \bar{P})^2} \sin \phi_2 \right].
\label{eq:av final_z}
\end{equation}

Let us now calculate the energy and its oscillatory part: $\varepsilon = \sqrt{m^2+P^2_x+P_y^2+(\bar{P}_z+\delta P_z)^2}$.
With Eqs.~(\ref{eq:av final_Px},\ref{eq:av final_Py}),
\begin{equation}
\varepsilon = \biggl[ m^2+m^2 \xi_1^2+m^2 \xi_2^2 + 2p_x \left( m \xi_1 \cos \phi_1 + m \xi_2 \cos \phi_2 \right) +\bar{P}_z^2  + p_x^2 +
2 \bar{P}_z \delta P_z+ 2 m^2 \xi_1 \xi_2 \cos(\phi_1-\phi_2) +\delta P_z^2 \biggr]^{1/2}.
\label{eq:av Emid}
\end{equation}
Using $\delta P_z$ given by Eq.~(\ref{eq:av final_Pz}) and recalling that $(k_1-k_2) \cdot \bar{P}=-2 \omega \bar{P}_z$, one can find out that the terms proportional to $ \cos (\phi_1-\phi_2)$ cancel each other. The expression for the energy may be further simplified to
\begin{equation}
\varepsilon = \Biggl\{ m^2+m^2 \xi_1^2+m^2 \xi_2^2
+p_x^2 + \bar{P}_z^2  + \delta P_z^2 +2p_x  \left[ \frac{m \xi_1}{ \left( 1- \bar{v}_z \right)} \cos \phi_1 + \frac{m \xi_2}{ \left( 1+ \bar{v}_z \right)} \cos \phi_2 \right]
 \Biggr\}^{1/2}
\end{equation}
with
$k_1 \cdot \bar{P} = \omega \bar{\varepsilon} \left( 1- \bar{v}_z \right)$, $ k_2 \cdot \bar{P} =  \omega \bar{\varepsilon} \left( 1+ \bar{v}_z \right)$,
\label{eq:av kdP}
where the average velocity on the $z$ axis is defined as $\bar{v}_z = \bar{P}_z / \bar{\varepsilon}$.
With a Taylor expansion the following expression is obtained
\begin{equation}
\varepsilon \approx \bar{\varepsilon} + \delta \varepsilon+ \bar{\varepsilon} \, O \left( \frac{\delta P_z}{\bar{\varepsilon}} \right)^2,
\label{eq:av E_taylor}
\end{equation}
where
the average energy, effective mass, and the oscillatory part are defined as
\begin{equation}
\bar{\varepsilon} = \sqrt{m_*^2+p_x^2+\bar{P}_z^2}, \quad m_* \equiv m \sqrt{1+\xi_1^2 + \xi_2^2}, \quad \delta \varepsilon= p_x \omega \left[ \frac{m \xi_1}{k_1 \cdot \bar{P}} \cos \phi_1 + \frac{m \xi_2}{k_2 \cdot \bar{P}} \cos \phi_2 \right].
\label{eq:av m_eff}
\end{equation}
Notice that for vanishing transverse momentum $p_x=0$ the energy is constant, in accordance with \cite{Avetissian_b_2016}.
The expansion in Eq.~(\ref{eq:av E_taylor}) is justified if $\delta P_z\ll \bar{\varepsilon}$.
Here we have taken into account that for an ultrarelativistic electron, the amplitude of $\delta P_z$ is always larger than $\delta \varepsilon$ according to Eqs.~(\ref{eq:av final_Pz}) and (\ref{eq:av m_eff}) and thus $\delta \varepsilon/\varepsilon < \delta P_z/\varepsilon \ll 1 $.
Taking into account the explicit form of $\delta P_z$, Eq.~(\ref{eq:av final_Pz}), the validity condition $\delta P_z\ll \bar{\varepsilon}$ reads
\begin{eqnarray}
\frac{m \xi_1 p_x \omega}{k_1 \cdot \bar{P}\bar{\varepsilon}} & =&\frac{p_x m \xi_1}{(1-\bar{v}_z)\bar{\varepsilon}^2}\ll 1\label{eq:av condP1}\\
  \frac{m \xi_2 p_x \omega}{k_2 \cdot \bar{P}\bar{\varepsilon}} & =&  \frac{p_x m \xi_2}{(1+\bar{v}_z)\bar{\varepsilon}^2}
\ll 1
\label{eq:av condP2}\\
\frac{ 2m^2 \xi_1 \xi_2 \omega}{ (k_1-k_2) \cdot \bar{P}\bar{\varepsilon}}
& =& \frac{m^2 \xi_1 \xi_2}{\bar{v}_z \bar{\varepsilon}^2}
\ll 1.
\label{eq:av condP3}
\end{eqnarray}

So far $\phi_1,\phi_2$ were not specified yet. With the help of $\delta \varepsilon,\delta P_z$ we evaluate $\delta \phi_1,\delta \phi_2$ and thus obtain the phases $\phi_1,\phi_2$.  Accordingly, the validity criterion for the basic assumption of this derivation, Eq.~(\ref{eq:key}), is determined.
Substituting Eqs.~(\ref{eq:av final_Pz}, \ref{eq:av m_eff}) in (\ref{eq:av delPhi}) we have
\begin{eqnarray}
\delta \phi_1= \Phi_1 + C_1 \sin \phi_2 - C_{12} \sin (\phi_1 - \phi_2),\quad
  \delta \phi_2 = \Phi_2+C_2 \sin \phi_1 + C_{12} \sin (\phi_1 - \phi_2),
\label{eq:av delphi2}
\end{eqnarray}
where $\Phi_1,\Phi_2$ are arbitrary constants and the coefficients are
\begin{eqnarray}
C_1  \equiv  \frac{2 p_x m \xi_2 \omega^2}{(k_2 \cdot \bar{P})^2}=
\frac{2 p_x m \xi_2 }{\varepsilon^2 (1+  \bar{v}_z)^2} \quad
C_2   \equiv  \frac{2 p_x m \xi_1 \omega^2}{(k_1 \cdot \bar{P})^2}=
\frac{2 p_x m \xi_1 }{\varepsilon^2 (1-  \bar{v}_z)^2}  \quad
C_{12}   \equiv \frac{ 2m^2 \xi_1 \xi_2 \omega^2}{[(k_1-k_2 ) \cdot \bar{P}]^2}
=\frac{ m^2 \xi_1 \xi_2  }{2 \varepsilon^2 \bar{v}^2_z} \,. \label{eq:av C_coef}
\end{eqnarray}
Eq.~(\ref{eq:av phi2Def}) together with
Eq.~(\ref{eq:av delphi2}) form an implicit system for the solution of the phases. Without loss of generality, we assumed that $\bar{v}_z>0$, i.e. the particle copropagates with the $\xi_1 $ beam, leading to asymmetry between the two beams. As a consequence, $k_1 \cdot \bar{P} \ll k_2 \cdot \bar{P}$, so that if the beams amplitudes are of the same order of magnitude, $C_2 $ is considerably larger than $C_1$.
In the following we assume that $C_1,C_{12} \ll 1$, yielding the following expressions
 \begin{eqnarray}
\phi_1(\tau)  \approx  \Phi_1+ \frac{k_1 \cdot \bar{P}}{m} \tau , \quad
\phi_2(\tau)    \approx \Phi_2+\frac{k_2 \cdot \bar{P}}{m} \tau  + \frac{2 p_x m \xi_1 \omega^2}{(k_1 \cdot \bar{P})^2} \sin \phi_1.
\label{eq:av phi2F}
\end{eqnarray}
In order to prove the consistency of this conclusion, one should accomplish two things. First, one has to show that the contributions of $C_1,C_{12}$ to the momentum are of second order, justifying the neglection. For this purpose, we consider a general function $F$ with the following argument $\phi(\tau)= \phi_0(\tau) + \nu \sin f(\tau)$, where $ \nu$ is a small constant and $\phi_0(\tau), f(\tau)$ are general functions. Taylor expanding with respect to $\nu$ yields
 \begin{equation}
F[\phi(\tau)] \approx
F[\phi_0(\tau) ]
+
\nu F'[\phi_0(\tau) ]  \sin f(\tau).
\label{eq:av identity0}
\end{equation}
In our case, $\phi_0$ designates the approximated phases $\phi_1$ or $\phi_2 $ given in Eq.~(\ref{eq:av phi2F})
 and $\nu $ is either $C_1$ or $C_{12}$, $\phi$ stands for the full phases including the neglected terms proportional to $C_{12},C_1$,  and $F(\phi)$ either $\epsilon \cos(\phi) $ or $\epsilon \sin \phi$, where $\epsilon$ stands for the amplitudes of the various momentum oscillations appearing in Eqs. (\ref{eq:av final_Px}), (\ref{eq:av final_Py}),(\ref{eq:av final_Pz}). Since $F' \sim \epsilon$, the correction scales as $O(\epsilon \nu)$. One should notice that the amplitude of the momentum oscillations are assumed to be considerably smaller with respect to the particle energy, being the dominant energy scale.
Hence, $\epsilon$ is a small parameter and the corrections corresponding to $C_1,C_{12}$ may be neglected, up to the second order.

Second, one should verify that the approximation of Eq.~(\ref{eq:key}) indeed holds.
Plugging the phases Eq.~(\ref{eq:av phi2F})
into Eq.~(\ref{eq:key}) we notice that all the three integrals take the form
$\mathcal{I} \equiv \int \cos \left[ \alpha \tau +  \beta \sin ( \kappa \tau) \right] d \tau$ with different choices of $\alpha,\beta,\kappa$.
In order to calculate this integral, we recall the identity
\begin{equation}
e^{i \beta \sin ( \kappa \tau)} = \sum_s J_s(\beta) e^{is \kappa \tau},
\label{eq:av identity1}
\end{equation}
where $J_s(\beta)$ is the Bessel function. Multiplying $e^{i \alpha \tau}$ on both sides, one readily obtains the real and imaginary part, respectively, as
\begin{equation}
\begin{aligned}
\cos \left[ \alpha \tau +  \beta \sin ( \kappa \tau) \right] &= \sum_s J_s(\beta) \cos \left[ (\alpha +s \kappa) \tau \right] \, ; \,
\sin \left[ \alpha \tau +  \beta \sin ( \kappa \tau) \right] &= \sum_s J_s(\beta) \sin \left[ (\alpha +s \kappa) \tau \right].
\end{aligned}
\label{eq:av identity2}
\end{equation}
The integral can thus be obtained as
\begin{equation}
\mathcal{I} = -\frac{1}{2} \sum_s J_s(\beta) \frac{1}{\alpha+s \kappa} \left[e^{i(\alpha+s \kappa) \tau} +e^{-i(\alpha+s \kappa) \tau } \right] \,.
\end{equation}
For a certain $\beta$ we know that $J_s(\beta)$ vanishes  if the index $s$ is  larger enough than $\beta$. Therefore, further simplification can be accomplished if
\begin{equation}
 s \kappa / \alpha \lesssim \beta \kappa / \alpha \ll 1.
\label{eq:av I1cond}
\end{equation}
The integral is thus approximated by
\begin{equation}
\mathcal{I} \approx -\frac{1}{\alpha} \cos \left[ \alpha \tau +  \beta \sin ( \kappa \tau) \right],
\label{eq:av I1approx}
\end{equation}
where Eq.~(\ref{eq:av identity2}) has been considered. This result is in agreement with Eq.~(\ref{eq:key}). Now let us find the conditions for which Eq.~(\ref{eq:av I1cond}) is satisfied for all three cases. For the first integral, $\beta$ vanishes and Eq.~(\ref{eq:av I1cond}) is trivially fulfilled. For the second case, one has $\alpha= (k_2 \cdot \bar{P})/m^2, \kappa= (k_1 \cdot \bar{P})/m^2$ and $\beta=C_{2}$, so that Eq.~(\ref{eq:av I1cond}) yields
\begin{equation}
\frac{p_x m \xi_1}{\bar{\varepsilon}^2} \ll \frac{(1+\bar{v}_z)(1+\bar{v}_z)}{2}.
\label{eq:av cond4}
\end{equation}
For the third integral $\beta, \kappa$ are as in the second case but $\alpha = [(k_2-k_1) \cdot \bar{P}]/m^2$, imposing the condition
\begin{equation}
\frac{p_x m \xi_1}{m_*^2} \ll \bar{v}_z (1-\bar{v}_z).
\label{eq:av cond5}
\end{equation}
Thus, Eq.~(\ref{eq:key}) was explicitly shown to be valid, given that Eqs.~(\ref{eq:av cond4}) and (\ref{eq:av cond5}) are satisfied.
Combining $C_1,C_{12} \ll 1$ with Eqs.~(\ref{eq:av condP1}-\ref{eq:av condP3}) and (\ref{eq:av cond4}-\ref{eq:av cond5}), yields the final validity criteria
\begin{eqnarray}
 \frac{m^2 \xi_1 \xi_2}{\varepsilon^2}   \ll  \bar{v}_z \min \left[ 1,2\bar{v}_z \right], \quad
 \frac{p_x m \xi_2}{\varepsilon^2}     \ll   (1+\bar{v}_z)^2,\quad
 \frac{p_x m \xi_1}{\varepsilon^2}      \ll   \bar{v}_z(1-\bar{v}_z).
\label{eq:av cirterion3F}
\end{eqnarray}
Let us conclude the derivation. The final expressions for the trajectory and momentum are Eqs.~(\ref{eq:av final_x},\ref{eq:av final_y},\ref{eq:av final_z}) and Eqs.~(\ref{eq:av final_Px},\ref{eq:av final_Py},\ref{eq:av final_Pz},\ref{eq:av m_eff}), correspondingly. The phases $\phi_1 (\tau)$ and $\phi_2(\tau)$ are given by Eq.~\eqref{eq:av phi2F}. The validity criteria corresponding to this solution are Eqs.~(\ref{eq:av cirterion3F}).
In the ultrarelativistic regime $1-\bar{v}_z \ll 1$ they are simplified to
\begin{equation}
 \frac{p_x m \xi_2}{2\varepsilon^2} ,
  \frac{m^2 \xi_1 \xi_2}{\varepsilon^2} ,
 \frac{2p_x m \xi_1}{m_*^2}  \ll 1.
\label{eq:av cirterionR}
\end{equation}
The above criteria can be fulfilled in a scenario where an ultrarelativistic electron moves along the laser propagation direction with a small deviating angle.
Alternatively, one may write the instantaneous momentum in a covariant form as follows
\begin{equation}
P_{\mu}(\tau)= \bar{P}_{\mu}-e\left[ A^{\mu}_1(\phi_1)+A^{\mu}_2(\phi_2) \right] + k^{\mu}_1 \left[
\frac{ep \cdot A_1(\phi_1)}{k_1 \cdot \bar{P}} - \frac{ A_1(\phi_2) \cdot A_2(\phi_2) }{ (k_1-k_2) \cdot \bar{P}}  \right] + k^{\mu}_2 \left[
\frac{ep \cdot A_2(\phi_2)}{k_2 \cdot \bar{P}} + \frac{ A_1(\phi_2) \cdot A_2(\phi_2) }{ (k_1-k_2) \cdot \bar{P}}  \right].
\label{eq:av final_LI}
\end{equation}
One may verify that in the case if one of the laser beams vanishes, our result Eq.~(\ref{eq:av final_LI}) recovers the familiar plane wave solution \cite{Ritus_1985}.

The above derivation expresses the physical quantities of interest, namely the trajectory and the 4-momentum, as a function of the proper time $\tau$. However, for practical applications it is favorable to use the laboratory time as the independent variable.
The two quantities are simply related through $dt=\frac{\varepsilon}{m}d \tau $.
Performing the integration we obtain
\begin{equation}
t(\tau)=   \frac{\bar{\varepsilon}}{m} \tau +  p_x \omega \left[ \frac{m \xi_1}{(k_1 \cdot \bar{P})^2} \sin \phi_1(\tau) + \frac{m \xi_2}{(k_2 \cdot \bar{P})^2} \sin \phi_2(\tau) \right].
\label{eq:av final_t}
\end{equation}
The latter along with $\textbf{x}(\tau)$ provides a parametric description of the particle coordinate as a function of the laboratory time.
Alternatively, one may further approximate the phases. We start by writing Eq.~(\ref{eq:av final_t}) as
\begin{equation}
\tau= \frac{m}{\bar{\varepsilon}} \biggl\{ t -
 p_x \omega \left[ \frac{m \xi_1}{(k_1 \cdot \bar{P})^2} \sin \phi_1(\tau) + \frac{m \xi_2}{(k_2 \cdot \bar{P})^2} \sin \phi_2(\tau) \right] \biggr\}.
\label{eq:av tau_mid}
\end{equation}
Substituting (\ref{eq:av tau_mid}) into the phase $\phi_1$ given in (\ref{eq:av phi2F}) one obtains
\begin{equation}
\phi_1 =  \Phi_1+\omega_1 t - p_x \omega \omega_1 \left[ \frac{m \xi_1}{(k_1 \cdot \bar{P})^2} \sin \phi_1 + \frac{m \xi_2}{(k_2 \cdot \bar{P})^2} \sin \phi_2 \right],
\label{eq:av phi1A}
\end{equation}
where $\omega_1 \equiv (1-\bar{v}_z) \omega$.
This equation is implicit, since $\phi_1$ appears in both sides.
Nevertheless, it proves useful as a starting point for approximation of the phases, as we immediately show.
According to the validity condition Eq.~(\ref{eq:av cirterion3F}), one notices that the coefficients of the sine functions in Eq.~(\ref{eq:av phi1A}) are much smaller than 1. As a result, Eq.~(\ref{eq:av identity0}) may be employed here. The fact that Eq.~(\ref{eq:av phi1A}) is implicit ($\phi_1$ appears in both sides) poses no difficulty, since the argument $f(\tau)$ in (\ref{eq:av identity0}) is general and has no influence on the final result.
Due to (\ref{eq:av identity0}) and according to the same reasoning that led us to neglect $C_1,C_{12}$, they may be omitted, leading to
\begin{eqnarray}
\phi_1(t)  \approx  \Phi_1+  \omega_1 t, \quad \phi_2(t)  \approx  \Phi_2+\omega_2 t  + \frac{2 p_x m \xi_1 \omega^2}{(k_1 \cdot \bar{P})^2} \sin (\omega_1 t),
\label{eq:av phi2F2}
\end{eqnarray}
where  $\omega_2 \equiv (1+\bar{v}_z) \omega$. Hence, one observes that $\omega_1,\omega_2$ are the characteristic oscillation frequencies associated with the $\xi_1,\xi_2$ beams, respectively. Notice that according to our convention the particle copropagates with the $\xi_1$ beam, so that $\bar{v}_z$ is positive, and hence $\omega_2$ is considerably  larger than $\omega_1$, which indicates the non-resonant regime of interaction.
\begin{figure*}
  \begin{center}
  \includegraphics[width=1\textwidth]{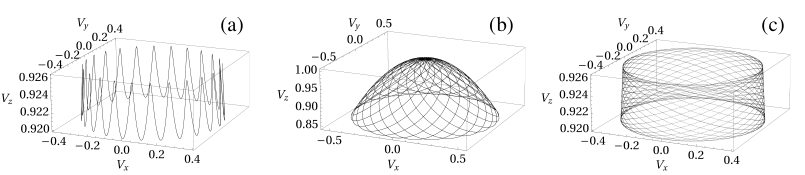}
  \caption{The electron velocity for three different field configurations. In all cases, the electron is copropagating with the $\xi_1$ laser beam and the velocity is shown for the time period during which the electron has travelled only for one cycle in the $\xi_1$ laser beam.  Simulation parameters are: (a) $\xi_1=50,\xi_2=1$; (b) $\xi_1=20,\xi_2=20$; (c) $\xi_1=1,\xi_2=50$. In all cases the electron has no transverse momentum and its energy in the field is $\varepsilon=2.6m_*$.}
  \label{fig:trajectory}
  \end{center}
\end{figure*}
\begin{figure*}[b]
  \begin{center}
  \includegraphics[width=0.9\textwidth]{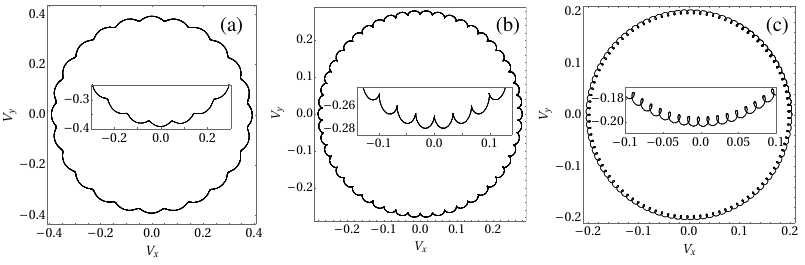}
  \caption{The transverse velocity in the $x-y$ plane for $\xi_1=50,\xi_2=1,p_x=0$ with three electron energies: (a) $\varepsilon=130m$; (b) $\varepsilon=182m$; (c) $\varepsilon=250m$. The initial transverse momentum in all panels is zero. The inset is just the zoom in of the velocity for a small time interval. The plots are for the time period during which the electron has travelled only for one cycle in the $\xi_1$ laser beam.}
  \label{fig:2D}
  \end{center}
\end{figure*}

\subsection{Characteristics of the trajectories}

With the obtained analytical expression for the electron momentum and coordinate, we  study in this section the main characteristics of the motion. As the dynamics is strongly effected by both of the laser beams,
we would expect to find some unusual features in the electron trajectory, where the acceleration is large and which may yield radiation emission deviating from the LCFA results based on the Baier-Katkov technique \cite{Baier_b_1994}.
We inspect the electron velocity in all components for 3 different field parameters in Fig.~\ref{fig:trajectory},
 featuring various behaviors.
 The results shown in the figure are obtained within the analytical treatment presented above and proved by the fully numerical solutions of Eq.~(\ref{eq:av EOM}).
 In all cases the initial transverse momentum vanishes $p_x=0$ and the energy is $\varepsilon=2.6m_*$, corresponding to $\omega_2/\omega_1 \approx 25$. The plots present a time interval of $2 \pi/\omega_1$, so it consists of one cycle of the $\xi_1$ beam and about 25 cycles of the $\xi_2$ beam.

In panel (a) the laser parameters are $\xi_1=50,\xi_2=1$. In the $x$-$y$ plane the particle performs a cyclic motion with a radius of $m \xi_1/\varepsilon$ and a frequency $\omega_1$ and on top of it rapid oscillations with frequency $\omega_2$ and amplitude $m \xi_2/\varepsilon$.
According to Eq.~(\ref{eq:av final_Pz}), the amplitude of the oscillation on the $z$ axis scales as $\sim m^2\xi_1 \xi_2 / \varepsilon^2$ and is, therefore, considerably smaller as compared to those in the $x,y$ axes.

Panel (b) depicts the case of $\xi_1=\xi_2=20$. In the $x$-$y$ plane the oscillations amplitude are now identical, so that the particle moves in circles with frequency $\omega_2$ according to $\xi_2$. An interesting point is that
the origin of the circle also exhibits a cyclic motion due to $\xi_1$ with a frequency of $\omega_1$. Both the fast $\xi_2$ circle and the slow $\xi_1$ circle have the same radius because of the identical oscillation amplitudes. In addition, one can observe that the tilting angle of the total velocity with respect to the $z$ axis is gradually changing. The reason is that the oscillation frequency on the $z$ axis is $\omega_2-\omega_1$. As a result, the relative phase between $v_z$ and $v_x$ for example gradually increases during the time interval under consideration from 0 to $2 \pi$.

Panel (c) presents the dynamics for $\xi_1=1,\xi_2=50$. It is quite similar to the previous case, but now the radius of the slow $\xi_1$ circle is negligible, such that the motion takes the form of a single circle with time dependent tilt.

With respect to radiation emission, the more irregular the trajectory is, the more interesting is the spectral shape. Hence, in the following we concentrate on the $\xi_1 \gg \xi_2$ case, like in panel (a) of Fig.~\ref{fig:trajectory}, where the dynamics is much more complex.  Fig.~\ref{fig:2D} shows a two dimensional projection of the velocity on the $x$-$y$ plane for $\xi_1=50,\xi_2=1$ with three different particle energies $\varepsilon$.

Panel (a) corresponds to $\varepsilon=130m$. As mentioned in Fig.~\ref{fig:trajectory}, one can see that the dynamics is a combination of a large circle due to $\xi_1$ and rapid oscillations corresponding to $\xi_2$, which have a smooth sine-shape, see in the inset. When the energy is increased, see panel (b) with $\varepsilon=182m$, several interesting changes take place. First of all, the number of the $\xi_2$ oscillations contained in one cycle of $\xi_1$ increases since the ratio of the frequencies, $\omega_2/\omega_1$, is now about $51$ instead of $25$ in panel (a). Furthermore, the radius of the circle as well as the amplitude of the small oscillations becomes smaller. This is because the amplitude of the transverse velocity $v_{\bot} \sim m\xi_1/\varepsilon$ decreases with the energy increase. More interestingly, a sharp spike-like feature emerges for each cycle of $\xi_2$ oscillation. It should be emphasized that the time scale corresponding to these spike-like features is significantly shorter than both $1/\omega_1$ and $1/\omega_2$.

In order to shed light on this spike-like feature, we take advantage of the approximated phases Eq.~(\ref{eq:av phi2F2})
and derive from the $y$ component of the trajectory (\ref{eq:av final_y}) the corresponding acceleration
\begin{equation}
\dot{v}_y =   \frac{m \xi_1 \omega_1}{\bar{\varepsilon}} \cos \phi_1 + \frac{m \xi_2 \omega_2 }{\bar{\varepsilon}} \cos \phi_2 = \frac{m^2}{\bar{\varepsilon}^2}\left( \chi_1\cos \phi_1 +  \chi_2   \cos \phi_2\right),
\label{eq:av vydot1}
\end{equation}
where in the last expression we take into account that the quantum parameter is proportional to the acceleration $\chi=\varepsilon^2|\dot{v}|/m^3$ \cite{Baier_b_1994}. Please note that here $\dot{a}$ refers to the derivative of time t. Here $\chi_1= \xi_1  \varepsilon \omega_1 /m$, and $\chi_2= \xi_2  \varepsilon \omega_2 /m$ are the quantum parameters induced, respectively, by beams 1 and 2. Let us take a close look at the time interval corresponding to $0<\phi_1<\pi/2$. One can see that as long as $\chi_2<\chi_1$, the acceleration does not change its sign. Namely, the velocity will monotonously decrease, as the case in Fig.~\ref{fig:2D}(a). Increasing the energy results in higher values of the ratio $\omega_2/\omega_1$, and at a certain point $\chi_2$ exceeds $\chi_1$. When $\chi_2$ becomes large enough, the acceleration $\dot{v}_y$ will change its sign during the time interval. If $\chi_2$ is only slightly higher than $\chi_1$, the acceleration is positive for a very short time, leading to sharp spikes, as encountered in Fig.~\ref{fig:2D}(b). In case $\chi_2$ is significantly larger than $\chi_1$, the acceleration is positive about half of the time, giving rise to the whirl appearing in Fig.~\ref{fig:2D}(c), where the energy is further increased to $\varepsilon=250m$. The impact of these phenomena on the radiation emission has been explored in Ref.~\cite{lv2021_anomalous}.

\begin{figure}
  \begin{center}
  \includegraphics[width=0.45\textwidth]{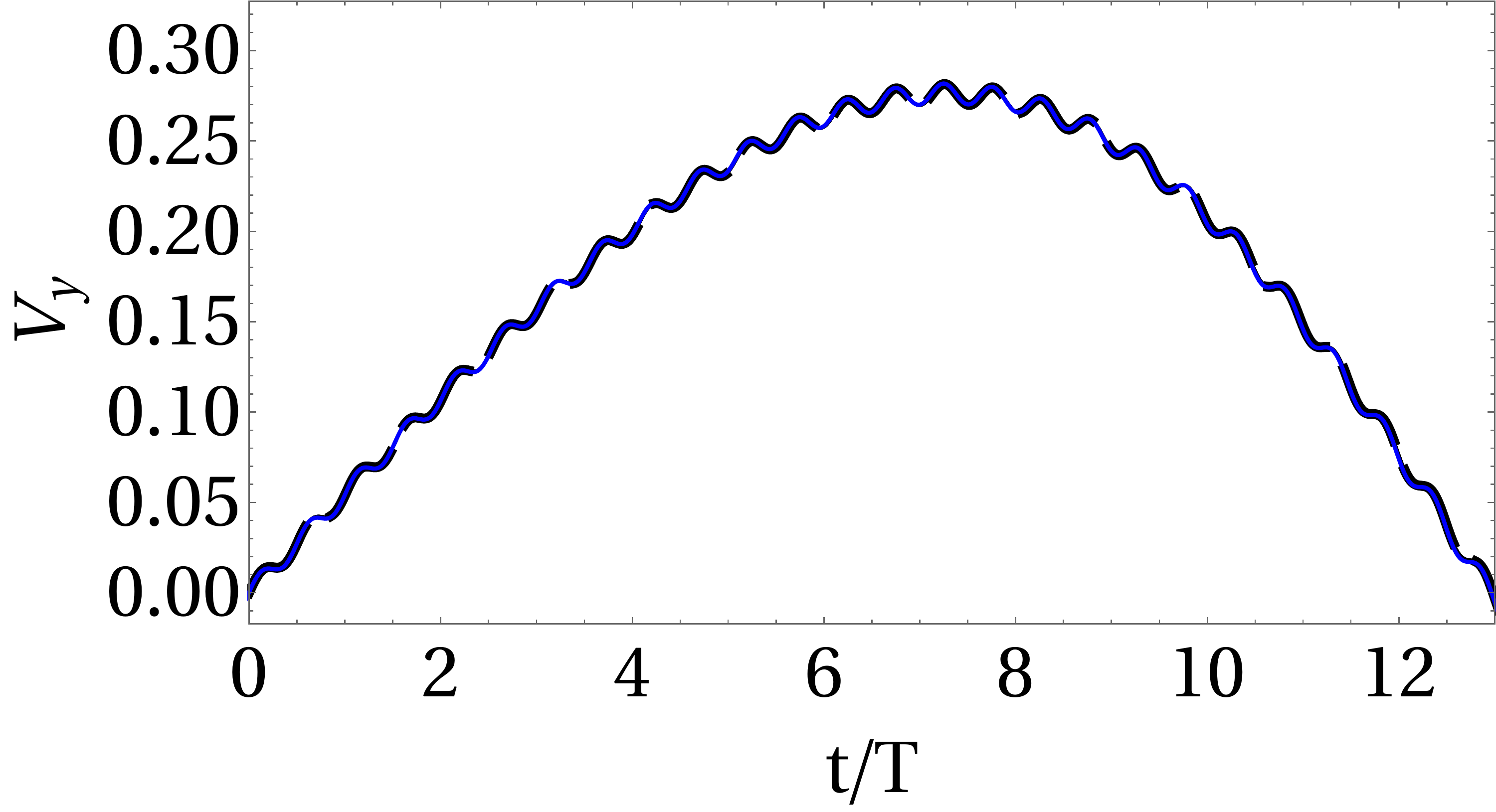}
  \caption{ The $y$ component of the velocity as a function of time is shown for the time duration in which the electron travels in $\xi_1$ laser beam for only half cycle. The electron has an average energy $\bar{\varepsilon}=182m$ copropagating with $\xi_1$ and the transverse momentum is $p_x=2.5m$. The field parameters are $\xi_1=50,\xi_2=1$. The time t is in units of $T=2 \pi/\omega$.}
  \label{fig:phaseMod}
  \end{center}
\end{figure}
Finally, let us examine the influence of the transverse momentum $p_x$. From the final expressions for the momentum and energy, one observes that this quantity has several contributions. First, it gives rise to the oscillations in the longitudinal momentum $P_z$ and the energy $\varepsilon$, see Eqs.~(\ref{eq:av final_Pz}) and (\ref{eq:av m_eff}), respectively. This means that the energy of the electron in the field is not constant anymore. Moreover, the non-zero transverse momentum also adds a slow sine-term (with frequency $\omega_1$) to the phase $\phi_2$, see Eq.~(\ref{eq:av phi2F}). As a result, the rapid oscillations corresponding to $\xi_2$ are periodically modulated. This phenomenon is demonstrated in Fig.~\ref{fig:phaseMod}, where the $y$-component of velocity is plotted as a function of time within half a cycle of $\xi_1$.  To verify our analytical results (black), the numerical solution is also shown in this figure as a blue line.
The agreement between the numerical solution and the analytical one is excellent, as the two curves are on top of each other. In addition, as we expected, the frequency of the small oscillations increases with time up to $t/T \approx 7$, with $T=2 \pi / \omega$, and then gradually decreases again.

\subsection{Drift momentum}
\label{sec:PzIN}

From the discussion above, we can see that the drift momentum of the particle in the laser fields, especially the average energy in the field, is an essential parameter for our approximation. However, the drift momentum depends on the asymptotic momentum of the particle before entering in the laser fields as well as the way of switching on the laser pulses. In this section, we will derive the relation explicitly.
The relation between $\bar{P}_{\mu}$ and the asymptotic momentum of the particle $p_{\mu}$ is governed by the ponderomotive force \cite{Bauer_1995,Mora_1998}, arising from the turn on process of the laser fields:
\begin{equation}
\frac{d \bar{P}_z}{d \tau} = -\frac{1}{2m } \frac{\partial}{\partial z} \overline{\left| e \textbf{A} \right| ^2}.
\label{pondeq}
\end{equation}
Substituting Eq.~\eqref{eq:av A_def} and keeping the envelope functions $g_1,g_2$, we have
\begin{equation}
\frac{d \bar{P}_z}{d \tau} = \frac{\omega m}{2} \left[ \xi_1^2 \frac{d}{d \phi_1} g_1^2[\phi_1(\tau)] - \xi_2^2 \frac{d}{d \phi_2} g_2^2[\phi_2(\tau)]
   \right].
\label{eq:av pond1}
\end{equation}
Suppose that the copropagating laser pulse with the amplitude $\xi_1$ is turned on first. During this process the second integral for $\bar{P}_z$ is vanishing, and the particle momentum reads 
\begin{equation}
 \bar{P}^{(1)}_z = p_z +
 \frac{\omega m \xi_1^2}{2} \int^{\tau}{d \tau'} \frac{d}{d \phi_1} g_1^2[\phi_1(\tau')]= p_z+ \frac{m^2 \xi_1^2 \omega}{2( k_1 \cdot p) } .
 \label{eq:av p1bar}
\end{equation}
Here $\phi_1 = (k \cdot p) \tau / m$, because in the absence of the counterpropagating pulse $k_1 \cdot P$ is exactly conserved. 
It is worthwhile to mention that this result is similar to the one corresponding to the plane wave case.
Now the second pulse is turned on. Its contribution to the momentum is given by
\begin{equation}
  \bar{P}_z =  \bar{P}^{(1)}_z -
 \frac{\omega m \xi_2^2}{2} \int^{\tau}{d \tau'}    \frac{d}{d \phi_2} g_2^2[\phi_2(\tau')].
 \label{eq:av pz_mid}
\end{equation}
 Recalling the approximation derived above Eq.~(\ref{eq:av identity0}), and assuming that the pulse is turned on adiabatically, namely $g_2'/g_2 \rightarrow 0$, the oscillatory part of the phase may be omitted, yielding
$g_2 (\phi_2) \approx g_2 \left( \frac{k_2 \cdot \bar{P}}{m} \tau \right)$.
 Since the first pulse effect comes into play through the neglected oscillatory term in $\phi_2$, it does not influence the integration. We further assume that $k_2 \cdot \bar{P}$ remains constant during the turn on of the second pulse.
Then, the integral in (\ref{eq:av pz_mid}) is straightforwardly carried out, yielding for $\bar{P}_z$ and  $\bar{\varepsilon}$:
\begin{equation}
 \bar{P}_z =p_z+ \frac{1}{2} \left[ \frac{m^2 \xi_1^2}{k_1 \cdot p}-\frac{m^2 \xi_2^2}{k_2 \cdot \bar{P}^{(1)}} \right] \omega,
 \quad\bar{\varepsilon} =p_0+ \frac{1}{2} \left[ \frac{m^2 \xi_1^2}{k_1 \cdot p}+\frac{m^2 \xi_2^2}{k_2 \cdot \bar{P}^{(1)}} \right] \omega,
\end{equation}
where Eqs.~(\ref{eq:av p1bar}), 
and $\bar{\varepsilon}=\sqrt{m_*^2 + \bar{P}_z^2}$ were employed.
Hence
\begin{equation}
 \bar{P}^{\mu} =p^{\mu}+  \frac{m^2 \xi_1^2}{2( k_1 \cdot p)} k_1^{\mu}+\frac{m^2 \xi_2^2}{2 [k_2 \cdot \bar{P}^{(1)}]} k_2^{\mu}.
\label{eq:av barPz}
\end{equation}
Examining the final momentum (\ref{eq:av barPz}), one may observe that our assumption $k_2 \cdot \bar{P} = k_2 \cdot \bar{P}^{(1)}$ was justified. We underline that
\begin{eqnarray}
k_1 \cdot \bar{P}= k_1 \cdot p + \frac{m^2 \xi_2^2 (k_1 \cdot k_2) }{2 [k_2 \cdot  \bar{P}^{(1)} ]},\quad
k_2 \cdot \bar{P}  = k_2 \cdot p + \frac{m^2 \xi_1^2 (k_1 \cdot k_2) }{2 [k_1 \cdot  p ]}.
\label{eq:av kdP2R}
\end{eqnarray}
Namely, neither $k_1 \cdot P$ nor $k_2 \cdot P$ are conserved. One may observe that $k_1 \cdot P$ is modified during the rise of the counterpropagating pulse and vice versa.
In case the counterpropagating beam is turned on first, an analogous derivation leads to
\begin{equation}
 \bar{P}^{\mu} =p^{\mu}+  \frac{m^2 \xi_1^2}{2[k_1 \cdot \bar{P}^{(2)}]} k_1^{\mu}+\frac{m^2 \xi_2^2}{2 (k_2 \cdot p)} k_2^{\mu},\label{eq:av barPz2}
\end{equation}
where $\bar{P}^{(2)}_{\mu} =p_{\mu}+ \frac{m^2 \xi_1^2}{2 (k_2 \cdot p )} k_{2\mu}$.
\begin{table}[b]
  \begin{center}
    \begin{tabular}{|c|c|c|c|c|c|c|}
      \hline
      case&$\xi_1$ & $\xi_2$ & $p_z$ &Order& $(\bar{\varepsilon},\bar{P}_z)^A$  &  $(\bar{\varepsilon},\bar{P}_z)^N$ \\
      \hline
       \multirow{2}{*}{1} & \multirow{2}{*}{10} & \multirow{2}{*}{10}& \multirow{2}{*}{0}  & a &(51.495,49.505) & (51.502,49.450) \\
        & & &   & b &(51.495,-49.505) & (51.502,-49.450) \\
      \hline
       \multirow{2}{*}{2} & \multirow{2}{*}{3} & \multirow{2}{*}{20}& \multirow{2}{*}{20}  &a& (200.637,199.613) & (200.637,199.612) \\
        &  & &   &b& (25.471,15.452) & (25.835,15.321) \\
      \hline
       \multirow{2}{*}{3} & \multirow{2}{*}{30} & \multirow{2}{*}{2}& \multirow{2}{*}{-1}  &a& (187.816,185.391) & (187.815,185.390) \\
        & &&  &b& (43.522,31.451) & (29.531,-0.396) \\
      \hline
    \end{tabular}
    \caption{The average 4-momentum of an electron after both laser beams are turned-on. We consider three different cases for different initial momentum $p_z$ and field parameters. In all the cases, the initial transverse momentum is chosen to be zero such that $\bar{P}_x=\bar{P}_y=0$. The fifth column, named Order, indicates the order by which the two laser beams are turned-on; (a) The $\xi_1$ beam is turned-on first. (b) The $\xi_2$ beam is turned-on first. The 4-momentum is given in units of the electron rest mass $m$. The superscript N designates the numerical calculation and A the analytical one.}
    \label{Tab:table1}
  \end{center}
\end{table}

The relation between the drift momentum and the asymptotic initial momentum has been also investigated by numerically solving the Lorentz equation (\ref{eq:av EOM}) and comparing with the analytical results. Table \ref{Tab:table1} presents the average 4-momentum of the electron after both laser beams are turned-on, corresponding to different initial momenta and intensities of the lasers. Since the order by which the lasers are turned-on affects the final state, the table contains both options. For the sake of simplicity, we assume the initial $p_{\bot}=0$ for all situations.
From the expression for the final momentum Eqs.~(\ref{eq:av barPz}) and (\ref{eq:av barPz2}), one can see that two factors determine which of the beams will be dominant. The obvious one is the corresponding field intensity. The surprising one is the relative direction between the propagation direction of the particle and the beam under consideration. It stems from the denominator $k \cdot \bar{P}$, namely, counterpropagating beams have lower influence than copropagating beams.

In the first case the two beams have identical intensity and the particle is initially at rest, so the only thing that breaks the symmetry is the turn-on order. It demonstrates that a given beam will have a stronger influence if it is the first to be turned on. The reason is that after the turned-on, the particle will copropagate with the first beam and thus this beam will have a large influence in the final results. This also reflects in the direction of the average momentum as in this case the particle always copropagates with the first beam at the end, see in Table \ref{Tab:table1}.


For the second case appearing in the table, one may naively assume the  $\xi_2$ beams should be dominant, since $\xi_2/\xi_1 \sim 7$ and the contribution to the final momentum of each beam scales like $\sim \xi^2$. However, due to the fact that $\xi_1$ is copropagating, its effect is actually of the same order of magnitude as of the $\xi_2$ beam.  This can be seen by the fact that the order of the turn-on causes  an order of magnitude difference between the final energies. Namely, when $\xi_1$ is turned-on first (Order (a)), the particle is first accelerated to ultrarelativistic energy and then slightly deccelerated  when $\xi_2$ is turned-on. The final energy is about $\varepsilon \approx 200m$, which is much larger than the final energy $\varepsilon \approx 25m$ of the second scenario (Order (b)), where the particle is first deccelerated and then accelerated.

In the third case a new situation is encountered. The particle flips its direction of motion during the turn-on of the second beam if the $\xi_2$ beam is turn-on first, Order (b) in the table. One may see that it initially propagates to the left and only after the second pulse rises it flips direction and propagates to the right. Both from analytical and from experimental perspectives, such a scenario should be avoided. From an experimental point of view, as it will lead to collisions of electrons in the beam with those following them. From analytical perspective, since a direction flip implies that the particle average velocity should vanish at a certain point in the middle of the turn-on process, violating the validity conditions. Indeed, the analytical expression in this case fails to reproduce the numerical result.

To complete the discussion, we specify several considerations which were taken into account when choosing the above parameters. First, we made sure that the validity criteria derived above are met. Second, the final propagation direction is always copropagating with the first turn-on beam, in agreement with the convention introduced in the previous subsection. Third, both laser amplitudes were chosen to be higher than 1. Since the contribution of each beam scales as $\propto \xi^2$, the influence of a beam with nonrelativistic intensity on the final momentum can be neglected.

\subsection{Systematic errors analysis of the trajectory}
\label{sec:erranaly}
\begin{figure}
  \begin{center}
  \includegraphics[width=0.4\textwidth]{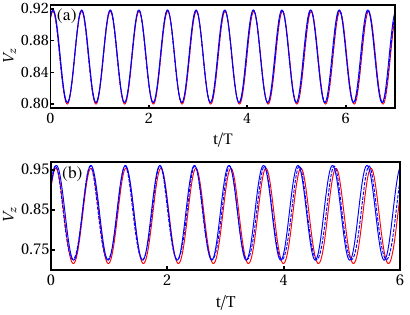}
  \caption{The analytical (blue) and numerical (red) longitudinal velocity for (a) $\epsilon_1=0.05$ and (b) $\epsilon_1=0.1$. In both cases $\epsilon_2=0$. The dashed line corresponds to advancing the analytical solution in time through the turn-on process and the solid line to determining $\Phi_1,\Phi_2$ according to the particle location when the turn on is over, see the main text. The time is in the unit of $T=2 \pi/\omega$.}
  \label{fig:error1}
  \end{center}
\end{figure}

\begin{figure*}
  \begin{center}
  \includegraphics[width=0.7\textwidth]{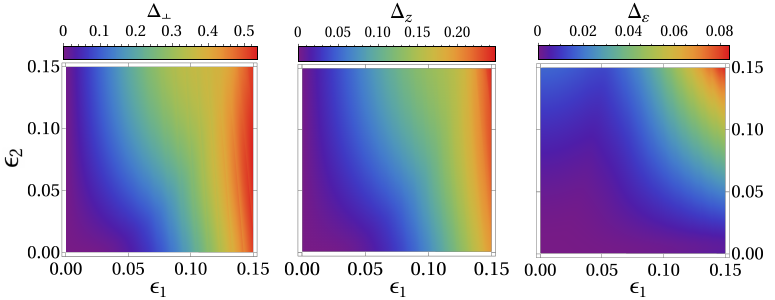}
  \caption{The relative deviation between analytical and numerical calculations of the following quantities: (a) the transverse velocity; (b) the longitudinal velocity ; (c) the energy. The $x,y$ axes are the small parameters stemming from the analytical derivation as $\epsilon_1 \equiv 2p_x m \xi_1/m_*^2$ and $\epsilon_2 \equiv m^2 \xi_1 \xi_2 / \bar{\varepsilon}^2$. The average energy is fixed to be $\varepsilon=200 m$ and the amplitude of the first beam is $\xi_1=100$. The initial momentum and the $\xi_2$ beam amplitude are in the range $0<p_x<15m$ and $0<\xi_2<60$, respectively.}
  \label{fig:error2}
  \end{center}
\end{figure*}

In the following, the accuracy of the analytical solution derived in the previous sections is systematically put to a test. For the sake of this purpose, we define the relative deviation of the analytical prediction (subscript $a$) of a quantity $X$ with respect to the numerically calculated value (subscript $n$) as follows
\begin{equation}
\Delta_X \equiv \frac{1}{2T}\int_{-T}^{T} dt \left| \frac{X_{a}-X_{n}}{X_{n}} \right|,
 \label{eq:av Delta_def}
\end{equation}
with $X$ being either the transverse velocity $v_{\bot} \equiv \sqrt{v_x^2 + v_y^2 }$, the longitudinal one $v_z$ or the energy $\varepsilon$.
The integration time is taken to infinity, i.e. $T \rightarrow \infty$. These deviations are explored as a function of $\epsilon_1 \equiv 2p_x m \xi_1/m_*^2$ and $\epsilon_2 \equiv m^2 \xi_1 \xi_2 / \bar{\varepsilon}^2$, being the small parameters of the derivation for ultrarelativistic particles, see Eq.~(\ref{eq:av cirterionR}). In the following we restrict ourselves to $\xi_1> \xi_2$, and hence the third small parameter is by definition smaller than $\epsilon_2$, namely $p_x m\xi_2/(2 \bar{\varepsilon}^2)<\epsilon_2$. For quantitative  comparison of the analytical and numerical quantities, we specify the arbitrary constants $\Phi_1,\Phi_2$ in the phases $\phi_1,\phi_2$, which is accomplished in two ways, see Fig.~\ref{fig:error1}. The first way (blue solid line) is to write down $\Phi_1=k_1 \cdot x_0, \Phi_2=k_2 \cdot x_0$ where $x_0$ denotes the temporal and spatial location of the particle at the moment when the turn-on process is over, associated with the numerical calculation. The second way (blue dashed line) is to advance the analytical solution in time from $t \rightarrow -\infty$ using the relation (\ref{eq:av barPz}). It should be mentioned that this approach is not fully analytical since $k_1 \cdot \bar{P} $ is not constant during the turn-on of the second pulse, but rather depends on $\xi_1$, which in its turn depends on $k_1 \cdot \bar{P}$ through the phase $\phi_1$. Nevertheless, assuming that the turn-on process is adiabatic, $\xi_1$ barely changes from one time step to its subsequent one, so that one can use the value of the field amplitude from the previous time step.

Before our broader parameter survey, it is worthwhile to take a close look at a specific case in order to gain intuition regarding the nature of deviation. Fig.~\ref{fig:error1} presents the longitudinal velocity for $\xi_1=50, \varepsilon=200m,p_x=0$ for a different $\xi_2$ value: $\xi_2=20$ and $\xi_2=40$ corresponding to $\epsilon_1=0.05$ (upper plot) and $\epsilon_1=0.1$ (lower plot), respectively. One can see that the main deviation stems from an inaccuracy in the phases $\phi_1,\phi_2$, rather than in the amplitudes.  
Moreover, the comparison between the dashed and solid curves in Fig.~\ref{fig:error1} demonstrates that the two approaches to determine the phases $\Phi_1,\Phi_2$ give similiar results and the first method was employed in the following calculations.

Fig.~\ref{fig:error2} depicts the relative deviations $\Delta_{\bot},\Delta_{z}, \Delta_\mathcal{\epsilon}$ defined in Eq.~(\ref{eq:av Delta_def}) as a function of $\epsilon_1,\epsilon_2$. We fixed the parameters $\xi_1=100, \varepsilon=200 m$ and varied $\xi_2,p_x$ in the ranges $0<\xi_2< 60$, $0<p_x<15 m$, respectively. The initial momentum on the $z$ axis was tuned in order to keep the energy $\varepsilon$ constant. One may notice that for vanishing $\xi_2$, the analytical calculation is accurate regardless of the value of $p_x$. This occurs due to the fact that vanishing of $\xi_2$ corresponds to the plane wave limit, where the analytical solution Eq.~(\ref{eq:av final_LI}) is exact without any restriction.

Furthermore, we can see from all three panels in Fig.~\ref{fig:error2} that the influence of the small parameter $\epsilon_2$ is considerably stronger as compared to that of $\epsilon_1$. This is because $\epsilon_2$ affects the phase while $\epsilon_1$ stems from the Taylor expansion of the energy (Eq.~(\ref{eq:av E_taylor})) and the discrepancy mainly originates from the dephasing in time, as shown in Fig.~\ref{fig:error1}. The amplitude of the oscillation in $v_\bot, v_z$ and $\varepsilon$, on the other hand, can be predicted quite well by the analytical expression even for nonnegligible $\epsilon_1,\epsilon_2$, which provides us a way to crudely estimate the relative error. For example, the average and oscillatory parts of $v_{\bot}$ may be roughly estimated, respectively, as $\sim m \xi_1/\varepsilon$ and $\sim m \xi_2  / \varepsilon$, and therefore the relative error is approximately $\Delta_\bot \sim \xi_2/\xi_1$. The relative error for $v_z$ is $ \sim m^2 \xi_1 \xi_2 / (\bar{v}_z\varepsilon^2)$, which is smaller than $\Delta_\bot$. For the energy, according to (\ref{eq:av m_eff}), the oscillations are closely related to $\epsilon_1$, i.e. $\delta \varepsilon/\varepsilon \approx \epsilon_1/2$. If we plug in the simulation parameters, these estimations qualitatively explain that for the same small parameter $\epsilon_1$ and $\epsilon_2$, $v_{\bot}$ has the largest deviation.
For Fig.~\ref{fig:error2}(c), we can see that the analytical results predict a very good approximation when $\epsilon_2=0$ no matter how large $\epsilon_1$ is. The reason is that when $p_x$ is zero the energy is constant (see Eq.~(\ref{eq:av m_eff})). Consequently, the phase plays no role and the approximation is quite good for the entire range of $\epsilon_1$ values presented in the figure.

\section{Radiation}
\label{sec:radcal}

Using the classical trajectory developed in Sec.~\ref{sec:clastra}, the radiation is calculated according to the Baier-Katkov method \cite{Baier_b_1994}.  
For the sake of simplicity, we start with a spinless particle.  Analogous derivation for the spinor case is given later.
The Baier-Katkov expression for the emitted intensity $I$ reads
\begin{equation}
dI = \frac{\alpha \varepsilon}{(2 \pi)^2 \varepsilon'T_0} |\mathcal{T}_{\mu}|^2 d^3\textbf{k}'
\label{eq:av BK}
\end{equation}
where $\alpha$ is the fine structure constant, $T_0$ is the interaction time, $\varepsilon' = \varepsilon - \omega'$, and
\begin{equation}
\mathcal{T}_{\mu}(k') = \int_{-\infty}^{\infty}{dt}v_{\mu}(t) e^{i \psi}, \quad \quad \psi \equiv \frac{\varepsilon}{\varepsilon'}k' \cdot x(t),
\label{eq:av T_def}
\end{equation}
where $v_{\mu}=dx_\mu / dt$, $k'_{\mu}$ is the emitted photon four-momentum
characterized by its energy $\omega'$ and the emission direction $\textbf{n}$ as
\begin{equation}
k'_{\mu} = \omega' \left(1, \textbf{n} \right) ,
\label{eq:av k_def}
\end{equation}
with
$\textbf{n} = \left( \cos \varphi \sin \theta, \sin \varphi \sin \theta, \cos \theta \right)$. In the realm of this theory, the oscillation $\delta \varepsilon$ are assumed to be small as compared to $\varepsilon$, which holds in our case as shown in the previous section. Accordingly, the factor appearing in the phase may be approximated as $\frac{\varepsilon}{\varepsilon'} \approx \frac{\bar{\varepsilon}}{\bar{\varepsilon}'}\left[ 1+\frac{(\delta \varepsilon)^2}{\bar{\varepsilon} \bar{\varepsilon}'} \right]$. In the following derivation the second order correction is neglected. Moreover, for simplicity reasons, the average energy $\bar{\varepsilon}$ is replaced from now on by $\varepsilon$. Furthermore, since the trajectory is given in terms of the proper time $\tau$, we change the integration variable in Eq.~(\ref{eq:av T_def}), leading to
\begin{equation}
\mathcal{T}_{\mu}(k') = \int_{-\infty}^{\infty}{d \tau} \frac{P_{\mu}(\tau)}{m} e^{i \psi}.
\label{eq:av T_tau}
\end{equation}

The actual calculation takes two steps. First, the Lorentz equation for a single particle is solved and the trajectory is obtained. Second, the time integration in the photon emission amplitude is calculated using the time-dependent momentum and coordinate of the particle. Both steps can be proceeded either analytically when an analytical trajectory is available or numerically for general laser field and electron beam parameters. In order to make sure that the integral in Eq.~\eqref{eq:av T_def} in our numerical calculation is converged, we have to solve the classical trajectory numerically with sufficiently small $\Delta t$ between two time steps, especially with large emitted photon energy as $\Delta t$ should be much smaller than $2\pi/\omega'$. For a high electron energy ($\varepsilon$ ranging from several hundreds of MeV to GeV) and a strong laser pulse (the classical parameter $\xi$ being around several hundreds) the analytical model is indispensable to calculate the whole spectrum, and numerically we confirm the result at discrete points in the spectrum. Below we will give an analytical derivation of the spectrum using the approximated trajectory of the electron given above. By substituting the trajectory and the emitted wavevector (\ref{eq:av k_def}) into expressions (\ref{eq:av T_tau}), the phase can be obtained like
\begin{equation}
\psi = \psi_{np} \tau -z^x_1\sin (\phi_1)+z^y_1\cos (\phi_1 ) -z^x_2\sin (\phi_1)+z^y_2\cos (\phi_2) -z_3\sin \left( \phi_1-\phi_2 \right) ,
\label{eq:av psi}
\end{equation}
where $u \equiv \frac{\omega'}{\varepsilon - \omega'}$, and the following quantities were introduced
\begin{equation}
z_1^x= \frac{m \xi_1 u}{\omega_1} \left[n_x + \frac{ p_x \omega }{\omega_1 \varepsilon} \left( n_z -1 \right) \right], \quad
z_1^y= \frac{m \xi_1 u}{\omega_1} n_y, \quad
z_2^x= \frac{m \xi_2 u}{\omega_2} \left[n_x - \frac{ p_x \omega }{\omega_2 \varepsilon} \left( n_z+1 \right) \right], \quad
z_2^y= \frac{m \xi_2 u}{\omega_2} n_y, \quad
z_3 = \frac{ m^2 \xi_1 \xi_2 u }{\bar{v}_z \Delta \omega \varepsilon} n_z,
\label{z3u}
\end{equation}
The linear term coefficient in (\ref{eq:av psi}) reads
\begin{equation}
\psi_{np} \equiv \frac{u \varepsilon^2}{m}   \left( 1 - \bar{v}_x \cos \varphi \sin \theta - \bar{v}_z \cos \theta \right) .
\label{psi_np}
\end{equation}
The phase $\psi$ may be simplified by introducing
\begin{equation}
\label{z1z2}
  z_1 = \sqrt{(z_1^x)^2 + (z_1^y)^2}, \quad
  z_2 = \sqrt{(z_2^x)^2 + (z_2^y)^2},\quad
  \varphi_1 =\tan^{-1} \left( \frac{z^y_1}{z^x_1}\right), \quad
  \varphi_2 =\tan^{-1} \left( \frac{z^y_2}{z^x_2}\right).
\end{equation}
Therefore, the phase takes the form
\begin{equation}
\psi = \psi_{np} \tau -z_1\sin (\phi_1-\varphi_1) -z_2\sin (\phi_1-\varphi_2) -z_3\sin \left( \phi_1-\phi_2 \right)   .
\label{eq:av psi_f}
\end{equation}
Notice that in the particular case of $p_x=0$, the second term in the expressions for $z_1^x, z_2^x$ vanishes, leading to
\begin{equation}
\label{eq:av z_px0}
z_1 = \frac{ m \xi_1 u \sin \theta}{\omega_1 },
\quad \quad
z_2 = \frac{ m \xi_2 u \sin \theta}{\omega_2 },
\quad \quad
z_3 \equiv \frac{ m^2 \xi_1 \xi_2 u }{\bar{v}_z \Delta \omega \varepsilon} \cos \theta
\end{equation}
as well as $\varphi_1=\varphi_2=\varphi$.
Let us calculate the $y$ component of $\mathcal{T}$ in detail. Employing (\ref{eq:av final_Py}) and (\ref{eq:av T_def}) we obtain
\begin{equation}
\mathcal{T}_{y} =     \int d \tau
\left[  \xi_1 \sin \phi_1 +  \xi_2 \sin \phi_2 \right] e^{i \psi}.
\label{eq:av T_y}
\end{equation}
In order to analytically solve this integral, the identity \cite{Ritus_1985}
\begin{equation}
(1,\cos \phi, \sin \phi) e^{-z \sin \left( \phi- \varphi \right)} = \sum_s (B_0,B_1,B_2) e^{-is \phi}.
\end{equation}
is invoked.
The functions $B_0,B_1,B_2$ are related to the Bessel function and its first derivative $J_s(z),J'_s(z)$ through
\begin{eqnarray}
B_0(s,z,\varphi) = J_s(z)e^{is \varphi}\quad
B_1(s,z,\varphi) = \left[ \frac{s}{z}J_s(z) \cos \varphi - i J'_s(z) \sin \varphi \right] e^{is \varphi}\quad
B_2(s,z,\varphi) = \left[ \frac{s}{z}J_s(z) \sin \varphi + i J'_s(z) \cos \varphi \right] e^{is \varphi} . 
\label{eq:av Bessel2}
\end{eqnarray}
As a result, the integral in (\ref{eq:av T_y}) is solved, yielding
\begin{equation}
\mathcal{T}_{y} = 2 \pi \sum_{s_1} \sum_{s_2}  \sum_{s_3}
 \delta(\Omega_{s_1,s_2,s_3}) \left[ \xi_1 B_0(\textbf{2})B_2(\textbf{1}) + \xi_2 B_0(\textbf{1})B_2(\textbf{2}) \right] B_0(\textbf{3}),
\end{equation}
where $\textbf{1} \equiv (s_1,z_1,\varphi)$, $\textbf{2} \equiv (s_2,z_2,\varphi)$ and $\textbf{3} \equiv (s_3,z_3,0)$ respectively, and the $\delta$ function argument is given by
\begin{equation}
\Omega_{s_1,s_2,s_3} \equiv \psi_{np} -\frac{\varepsilon}{m} \left[ s_1 \omega_1+ s_2 \omega_2 + s_3 \left( \omega_1 - \omega_2 \right) \right] \,.
\label{Om_rl}
\end{equation}
One may notice that different combinations of the indices $s_1,s_2,s_3$ may yield the same $\delta$ function argument. As a result, when squaring $\mathcal{T}$, interference terms will arise.
This interference depends on the quantity $\omega_2/\omega_1$. If this ratio is a rational number, the motion is periodic with the frequency $2 \pi/(n \omega_1)$ with $n$ being the decimal part of the rational number. Otherwise, the motion is non-periodic. In the following we discuss each of the cases separately.


\subsection{The non-periodic case}
\label{sec:nonperiodic}

In the non-periodic case, when the ratio $\omega_2/\omega_1$ is not an integer, it is convenient to define
$s_L \equiv s_1+s_3, s_R \equiv s_2-s_3$. Hence, one may write
\begin{equation}
\mathcal{T}_{y} = 2 \pi \sum_{s_L} \sum_{s_R} \mathcal{M}_{y}  \delta(\Omega_{s_L,s_R}),
\end{equation}
with $\Omega_{s_L,s_R} \equiv \psi_{np} - \frac{\varepsilon}{m} \left( s_L \omega_1 + s_R \omega_2 \right)$.
The matrix element takes the form
\begin{equation}
\mathcal{M}_y =   \sum_{s_3}  B_0(\textbf{3}) \left[ \xi_1 B_0(\textbf{2})B_2(\textbf{1}) + \xi_2 B_0(\textbf{1})B_2(\textbf{2}) \right] .
\label{eq:av My np}
\end{equation}
An analogous procedure may be applied for the other components as well, yielding
\begin{equation}
\mathcal{T}_{\mu} = 2 \pi \sum_{s_L} \sum_{s_R} \mathcal{M}_{\mu} (s_L,s_R,\omega',\cos \theta) \delta(\Omega_{s_L,s_R}),
\end{equation}
where
\begin{eqnarray}
\mathcal{M}_t &=&  \sum_{s_3}  B_0(\textbf{3}) \Biggl[ \frac{\varepsilon}{m} B_0(\textbf{1}) B_0(\textbf{2}) + \frac{p_x \omega \xi_1}{\omega_1 \varepsilon} B_1(\textbf{1}) B_0(\textbf{2}) + \frac{p_x \omega \xi_2}{\omega_2 \varepsilon} B_0(\textbf{1}) B_1(\textbf{2}) \Biggr],\label{eq:av Mt np} \\
\mathcal{M}_x &=& \sum_{s_3} \Biggl( \frac{p_x}{m} B_0(\textbf{1}) B_0(\textbf{2}) B_0(\textbf{3}) + B_0(\textbf{3}) \left[ \xi_1 B_0(\textbf{2})B_1(\textbf{1}) + \xi_2 B_0(\textbf{1})B_1(\textbf{2}) \right] \Biggr), \label{eq:av Mx np} \\
\mathcal{M}_z &=& \sum_{s_3}  \Biggl( B_0(\textbf{1}) B_0(\textbf{2}) \left[ \frac{\bar{P}_z}{m} B_0(\textbf{3}) - \frac{ m \xi_1 \xi_2 }{ \bar{v}_z \varepsilon} B_1(\textbf{3}) \right] + B_0(\textbf{3})\frac{p_x \omega }{ \varepsilon} \left[ \frac{ \xi_1 }{\omega_1 } B_1(\textbf{1}) B_0(\textbf{2})
-\frac{ \xi_2  }{\omega_2 } B_0(\textbf{1}) B_1(\textbf{2}) \right] \Biggr). \label{eq:av Mz np}
\end{eqnarray}
As the squaring $\mathcal{T} $ does not mix terms associated with different $s_L,s_R$ indices, the interference takes place only between terms included within $\mathcal{M}(s_L,s_R)$.
Finally, the emitted intensity may be obtained by integrating (\ref{eq:av BK}) over the polar angle.
\begin{equation}
\frac{dI}{d \omega' d \varphi} =
\frac{\alpha m}{ 2 \pi \varepsilon'}
 \int d(\cos \theta) \omega'^2 \sum_{s_L} \sum_{s_R} \Bigl| \mathcal{M}(s_L,s_R,\omega',\cos \theta)\Bigr|^2 \delta(\Omega_{s_L,s_R})
\label{eq:av emis1}
\end{equation}
where the identity $\delta^2(\Omega_{s_L,s_R})=\frac{\tau_0}{2 \pi}\delta(\Omega_{s_L,s_R})$ is used. The proper interaction time is given by $\tau_0 = (m/\varepsilon) T_0$.
The condition imposed by the $\delta $ function,  $\Omega_{s_L,s_R} = 0$,  determines the relation between $\cos \theta$ and $\omega',\varphi$
\begin{equation}
1-\rho - \bar{v}_z \cos \theta = \bar{v}_x \cos \varphi \sqrt{1-\cos^2 \theta}.
\label{eq:av cos1}
\end{equation}
Squaring and solving this equation one obtains two possible angles
\begin{equation}
\cos \theta_{\pm} = \frac{\bar{v}_z (1-\rho) \pm \bar{v}_x \cos \varphi \sqrt{\Delta}}{\bar{v}_z^2+\bar{v}_x^2 \cos^2 \varphi},
\label{eq:av cos2}
\end{equation}
where the following quantities were introduced
\begin{eqnarray}
\Delta   \equiv   \bar{v}_z^2+\bar{v}_x^2 \cos^2 \varphi - \left(1-\rho \right)^2, \quad
\rho   \equiv  \frac{\varepsilon'}{\varepsilon \omega'}
\left(
s_L \omega_1 + s_R \omega_2 \right).
\label{rho_def}
\end{eqnarray}
Notice that when squaring (\ref{eq:av cos1}) a redundant solution may be added, which solves the equation
\begin{equation}
1-\rho - \bar{v}_z \cos \theta =- \bar{v}_x \cos \varphi \sqrt{1-\cos^2 \theta},
\label{eq:av cos11}
\end{equation}
rather than the original one. Thus, the solutions given in (\ref{eq:av cos2}) are physical only when a positive results appear after substituting it into the right wing of (\ref{eq:av cos1}).
In quantitative terms, this condition reads
\begin{equation}
\frac{1-\rho}{\bar{v}_z} < \cos \theta \leq 1 .
\end{equation}
A solution that does not meet this criterion is therefore excluded.
Employing the $\delta$ function to perform the integration leads to
\begin{equation}
\frac{dI}{d \omega' d \varphi} =
\frac{\alpha \varepsilon \omega'^2}{ 2 \pi \varepsilon'}
  \sum_{i=\pm} \sum_{s_L} \sum_{s_R} \Bigl| \mathcal{M}(s_L,s_R,\omega',\theta_i) \Bigr|^2 \Bigl|\frac{d \Omega_{s_L,s_R}}{d(\cos \theta)} \Bigr|^{-1}_{\theta=\theta_i}.
\label{dIdphi}
\end{equation}
The reciprocal of the derivative of the $\delta$ function, required for the integration, reads
\begin{equation}
\Bigl|
\frac{d \Omega_{s_L,s_R}}{d(\cos \theta)} \Bigr|^{-1}
= 	\frac{m\varepsilon'}{\varepsilon^2  \omega'} \kappa \quad  \mathrm{with} \quad
\kappa \equiv
\left| \frac{1}{ \bar{v}_x \cos \varphi \cot \theta -\bar{v}_z } \right|.
\label{deriva}
\end{equation}
Plugging (\ref{deriva}) into (\ref{dIdphi}) the final result follows
\begin{equation}
\frac{dI}{d \omega' d \varphi} =
 \frac{\alpha \omega'm^2}{ 2 \pi \varepsilon^2} \sum_{i=\pm} \sum_{s_L} \sum_{s_R} \Bigl| \mathcal{M}(s_L,s_R,\omega',\theta_i) \Bigr|^2 \kappa(\theta_i) .
\label{eq:av FinalG}
\end{equation}

For spinor particle the initial emission expression (\ref{eq:av BK}) is modified as follows
\begin{equation}
|\mathcal{T}_{\mu}|^2 \rightarrow
|\mathcal{K}|^2 \equiv
-\left(
\frac{\varepsilon'^2 + \varepsilon^2 }{2 \varepsilon \varepsilon'}
 \right)
|\mathcal{T}_{\mu}|^2 +
\frac{\omega'^2}{2 \varepsilon'^2 \varepsilon'^2}
|\mathcal{T}_{0}|^2.
\end{equation}
Therefore, the final results for scalars (\ref{eq:av FinalG}) is multiplied by $\left(
\frac{\varepsilon'^2 + \varepsilon^2 }{2 \varepsilon \varepsilon'}
 \right)
$ and a second term is added
\begin{equation}
\frac{dI}{d \omega' d \varphi} = \frac{\alpha m^2 \omega'}{4 \pi \varepsilon^5 \varepsilon'}
\sum_{i=\pm} \sum_{s_L} \sum_{s_R} \kappa(\theta_i)  \left[
-\varepsilon^2 \left( \varepsilon^2 + \varepsilon'^2 \right)|\mathcal{M}_{\mu}(s_L,s_R)|^2  +
\omega'^2 m^2 |\mathcal{M}_0(s_L,s_R)|^2
\right].
\label{finalC}
\end{equation}

\subsection{The periodic case}
\label{sec:period}

Now, we consider the case of periodic motion when $\omega_2=n \omega_1$, with an integer $n$. As a result, the kinematic relation which follows from the $\delta$ function is modified. Using the relation between $s_L,s_R$ and $s_1,s_2,s_3$ one obtains $ s_L \omega_1 + s_R \omega_2 \rightarrow s_* \omega_1$, with the definition $ s_* \equiv s_1+ns_2-s_3(n-1)$. Accordingly, $\rho$ in (\ref{rho_def}) is replaced by $\rho = \frac{\varepsilon'}{\varepsilon \omega'} s_* \omega_1$. As a consequence of the periodicity, the summation over $s_2$ takes place inside the matrix element, similarly to $s_3$. Correspondingly, we have
\begin{eqnarray}
\mathcal{M}_t &=&  \sum_{s_2} \sum_{s_3}  B_0(\textbf{3}) \Biggl[ \frac{\varepsilon}{m} B_0(\textbf{1}) B_0(\textbf{2}) + \frac{p_x \omega \xi_1}{\omega_1 \varepsilon} B_1(\textbf{1}) B_0(\textbf{2}) + \frac{p_x \omega \xi_2}{\omega_2 \varepsilon} B_0(\textbf{1}) B_1(\textbf{2}) \Biggr],\label{eq:av Mt p} \\
\mathcal{M}_x &=& \sum_{s_2} \sum_{s_3} \Biggl( \frac{p_x}{m} B_0(\textbf{1}) B_0(\textbf{2}) B_0(\textbf{3}) + B_0(\textbf{3})
\left[ \xi_1 B_0(\textbf{2})B_1(\textbf{1}) + \xi_2 B_0(\textbf{1})B_1(\textbf{2}) \right] \Biggr), \label{eq:av Mx p} \\
\mathcal{M}_y &=& \sum_{s_2} \sum_{s_3}  B_0(\textbf{3}) \left[ \xi_1 B_0(\textbf{2})B_2(\textbf{1}) + \xi_2 B_0(\textbf{1})B_2(\textbf{2}) \right] ,
\label{eq:av My p} \\
\mathcal{M}_z &=& \sum_{s_2} \sum_{s_3}  \Biggl( B_0(\textbf{1}) B_0(\textbf{2}) \left[ \frac{\bar{P}_z}{m} B_0(\textbf{3}) - \frac{ m \xi_1 \xi_2 }{ \bar{v}_z \varepsilon} B_1(\textbf{3}) \right] + B_0(\textbf{3})\frac{p_x \omega }{ \varepsilon} \left[
\frac{ \xi_1 }{\omega_1 } B_1(\textbf{1}) B_0(\textbf{2}) -\frac{ \xi_2  }{\omega_2 } B_0(\textbf{1}) B_1(\textbf{2}) \right] \Biggr).\label{eq:av Mz p}
\end{eqnarray}
Compared to the non-periodic case, the interference between different harmonics in the spectrum is much more complicated in the periodic case as there is a double summation inside the squaring of the matrix elements. The final result, analogous to (\ref{finalC}) of the non-periodic case, is given by
\begin{equation}
\frac{dI}{d \omega' d \varphi} = \frac{\alpha m^2 \omega'}{4 \pi \varepsilon^5 \varepsilon'}
\sum_{i=\pm} \sum_{s_*} \kappa(\theta_i)  \left[
-\varepsilon^2 \left( \varepsilon^2 + \varepsilon'^2 \right)|\mathcal{M}^{\mu}_{s_*}|^2  +
\omega'^2 m^2 |\mathcal{M}^0_{s_*}|^2
\right].
\label{finalCP}
\end{equation}
It is worth to point out that the periodic case is most likely to be observed in a short laser pulse, when the condition $\omega_2= n \omega_1$ can be fulfilled within the broad bandwidth of the laser pulse. We discuss this issue below.

\subsection{Vanishing initial transverse momentum}
\label{sec:vanishing _px}
In this subsection several quantities are explicitly evaluated for the particular case of vanishing initial transverse momentum, $p_x=0$. It allows us to simplify the expressions and thus to obtain order of magnitude estimations which will prove useful later on.
Substituting $p_x=0$ to  (\ref{eq:av cos2}), the emitted photon angle reduces to
\begin{equation}
\cos \theta = \frac{1-\rho}{\bar{v}_z}
\end{equation}
Since $\rho \ll 1$, the corresponding sine function is approximately given by $\sin \theta \approx \sqrt{ 1 - \frac{1}{\bar{v}_z^2} + \frac{2 \rho}{\bar{v}_z^2}}$. Substituting this expression to the Bessel arguments definitions (\ref{eq:av z_px0}) one obtains
\begin{equation}
 \label{eq:av z1_i}
z_1 = \frac{\xi_1 m_*m}{\omega_1 \varepsilon} \sqrt{u(u_{s}-u)}, \quad z_2 = \frac{\xi_2 m_*m}{\omega_2 \varepsilon} \sqrt{u(u_{s}-u)}
\end{equation}
where we have defined $ u_{s} \equiv \frac{2 \varepsilon  \left( s_L \omega_1 + s_R \omega_2 \right)}{m_*^2}$ and the relation $ \bar{v}_z^2 = 1 - \frac{m^2_*}{\varepsilon^2} $ was employed. The maximal value of $z_1,z_2$ corresponds to $u = u_{s}/2$, namely
\begin{equation}
\label{eq:av Zmax}
z^{max}_{1} = \frac{ \xi_1 \left( s_L \omega_1 + s_R \omega_2 \right)}{\omega_1 m_*}= \frac{ m_* u \xi_1}{\omega_1 \varepsilon}, \quad
z^{max}_{2} = \frac{ \xi_2 \left( s_L \omega_1 + s_R \omega_2 \right)}{\omega_2 m_*}\frac{ m_* u \xi_2}{\omega_2 \varepsilon}.
\end{equation}

\subsection{Spectra in the strong field regime: $\xi_1 \gg 1$}
\label{sec:strongField}

In what follows we consider in detail the case where the copropagating beam is of relativistic intensity. It should be stressed that the spectrum may not be approximated by LCFA even though $\xi_1 \gg 1$. The physical conditions and the nature of this specific LCFA violation is discussed in \cite{lv2021_anomalous}.

In the strong field regime the argument of the Bessel function in Eq.~(\ref{z1z2}) can be the order of $10^8$ or even larger with the increasing of the laser field strength. This means the sum over the harmonics in the emission spectrum covers an extremely large region. In order to make the calculation feasible, we have employed an optimised scheme for the calculation, based on the logic proposed by Ritus \cite{Ritus_1985}.

It is well known that
an ultrarelativistic particle emits mainly within a cone of angle $\sim 1/\gamma$ along its propagation direction.
Hence, the emission angle $\theta$ may be approximated by the angle of the particle's momentum between $\textbf{P}$ with respect to the $z$ axis. Examining the classical momentum $\textbf{P}$, one observes that this angle lies in the range $\sin \theta_d < \sin \theta < \sin \theta_u$ and its time-averaged value is $\sin \theta_c$, where
\begin{eqnarray}
\sin \theta_c   \equiv   \frac{\sqrt{m_*^2+p_x^2}}{\varepsilon}, \quad
\sin \theta_d   \equiv   \frac{p_x \cos \varphi + m(\xi_1-\xi_2)}{\varepsilon}, \quad
\sin \theta_u  \equiv    \frac{p_x \cos \varphi + m(\xi_1+\xi_2)}{\varepsilon} \,.
\end{eqnarray}
In the case considered here, namely $\xi \gg 1, \xi_2 \ll \xi_1$, and due to $p_x \ll m\xi_1$ (see Eq. (\ref{eq:av condP1})), this range is very narrow and the angle may be crudely estimated according to the average value $\theta_c$.
Accordingly, one may show that the second term in the brackets appearing in the expression for $z_1^x,z_2^x$ is negligible. As a result, the $p_x=0$ expressions (\ref{eq:av z_px0}) provides an order of magnitude estimation for $z_1,z_2,z_3$. Plugging in $\sin \theta_c \approx m_*/\varepsilon, \cos \theta_c \approx 1$ one obtains
\begin{equation}
z^c_1 = \frac{m_* \xi_1 u}{\omega_1 \varepsilon}, \quad
z^c_2 = \frac{m_* \xi_2 u}{\omega_2 \varepsilon}, \quad
z^c_3 = \frac{m \xi_1 \xi_2 u}{\bar{v}_z \omega_2 \varepsilon}
 \label{eq:av Zc}
\end{equation}
Notice that $z_1^c,z_2^c$ coincide with the maximal value possible for these quantities, see Eq. (\ref{eq:av Zmax}).
Furthermore,  one may observe that since $\bar{v}_z \approx 1$ and $m_* \approx m \xi_1$ we have $z^c_3 \approx z^c_2$.

In the following we take advantage of these relations in order to accelerate the harmonics summation ($s_L,s_R,s_3$) appearing in the final emission formula (\ref{finalC}) as well as derive simplified validity conditions.
We follow the logic presented by Ritus \cite{Ritus_1985} for emission in a circularly polarized laser. Since $u \sim \chi$, these arguments may be much larger than 1. As a result, the number of harmonics contributing to the emission may be enormous, and an efficient way to carry out the summation is required. First, we replace the summation by integration. Second, since Bessel function of high order is maximal for $z \approx s$ and strongly suppressed for either $z \gg s$ or $z \ll s$, the integration is centred around
\begin{equation}
s^c_{L} = z^c_1+z^c_3, \quad
s^c_{R} = z^c_2 - z^c_3, \quad
s^c_{3}=z^c_3.
\end{equation}
In order to estimate the integration range, we define $z^d_1,z^d_2,z^d_3$ and $z^u_1,z^u_2,z^u_3$ analogously to $z^c_1,z^c_2,z^c_3$ appearing in Eq.~\eqref{eq:av Zc} with $\theta = \theta_d$ and $\theta = \theta_u$ respectively.
Accordingly, the upper and lower limits of the integration are respectively
\begin{equation}
s^u_{L} = z^u_1+z^u_3, \quad
s^u_{R} = z^u_2 - z^u_3, \quad
s^u_{3}=z^u_3.\quad
s^d_{L} = z^d_1+z^d_3, \quad
s^d_{R} = z^d_2 - z^d_3, \quad
s^d_{3}=z^d_3.
\end{equation}
In mathematical terms, our improved summation scheme may be formulated as
\begin{equation}
\sum_{s_L} \rightarrow \int_{s^d_L}^{s^u_L} d s_L\,, \sum_{s_R} \rightarrow \int_{s^d_R}^{s^u_R} d s_R\,, \sum_{s_3} \rightarrow \int_{s^d_3}^{s^u_3} d s_3
\label{eq:av acc_scheme}
\end{equation}
The replacement of the summation by integral in the calculation is appropriate only when $s_u-s_d$ is large enough, which strongly depends on the chosen parameters.

\begin{figure*}
  \begin{center}
  \includegraphics[width=0.8\textwidth]{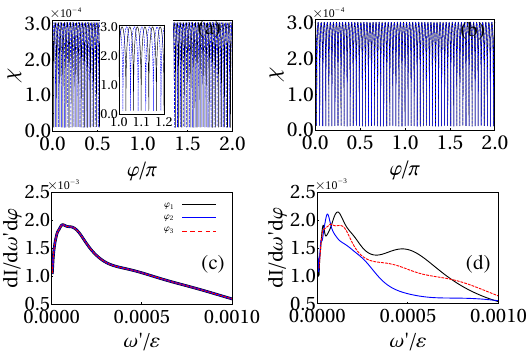}
  \caption{The time-dependent quantum parameter $\chi(t)$ as a function of $\varphi$ for both non-periodic case (a) and periodic case (b). The solid black curves are for the one cycle of the laser pulse and the dashed blue curves are for another cycle that is ten period away. The emission spectra for  non-periodic case (c) and periodic case (d). Here $\varphi_1$ is corresponding to $\chi(t)$ near maximum, $\varphi_2$ is for $\chi(t)$ in the middle and $\varphi_3$ is for $\chi(t)$ near minimum. The other parameters are $\xi_1=20$ and $\xi_2=0.3$ with $p_x=0$.  The non-periodic case is for $\varepsilon=4m_*$ with $\omega_2/\omega_1 \approx 60.1$, while the periodic case is for $\omega_2/\omega_1=60$ ($\varepsilon=4.02m_*$). }
  \label{fig:p_np_chi_spec}
  \end{center}
\end{figure*}

In the following we present typical spectra in the strong field regime ($\xi_1=20,\xi_2=0.3,p_x=0$), and use it to discuss the differences between the periodic and non-periodic cases derived above. Please note that the radiation reaction is neglected in the applied parameter regime, as the energy emitted during one laser cycle is very small compared with the electron energy. 

\textit{Non-periodic case versus periodic case with $p_x = 0$.} 
In Fig.~\ref{fig:p_np_chi_spec}(a) and (c), we consider the non-periodic case. The energy was chosen to be $\varepsilon=4 m_*$, so that the ratio $\omega_1/\omega_2$ is about 60.1.
Since the radiation  of an ultrarelativistic electron emitted to a certain direction originates from the vicinity of the location where the particle velocity points to the detector, it implies that the emission should depend on $\varphi$. However, due to the non-periodicity, we can see from  Fig.~\ref{fig:p_np_chi_spec}(a) that the emission at a given $\varphi$ takes place with
different $\chi$ values in different $\omega_1$ cycles.
As a result of this $\chi$-averaging, the difference between emissions at various $\varphi$ disappears
with long enough pulse of $\xi_1$, see the spectra in Fig.~\ref{fig:p_np_chi_spec}(c) for three different $\varphi$. The spectrum was evaluated with the aid of the non-periodic formula (\ref{finalC}) together with (\ref{eq:av acc_scheme}).

In the periodic case, Fig.~\ref{fig:p_np_chi_spec}(b) and (d), the electron energy is tuned $\varepsilon=4.02m_*$ to fulfil the integer ratio $\omega_2/\omega_1=60$. As opposed to the non-periodic case, here a particular value of $\chi$ parameter  corresponds to the emission at a given $\varphi$ at any period of the trajectory
(see Fig.~\ref{fig:p_np_chi_spec} (b)), and therefore the emission depends on $\varphi$. In Fig.~\ref{fig:p_np_chi_spec}(d) the black, blue and red curves are calculated with different values of $\varphi$, respectively. One may see that these three curves significantly differ from each other.

\begin{figure}[t]
  \begin{center}
  \includegraphics[width=0.5\textwidth]{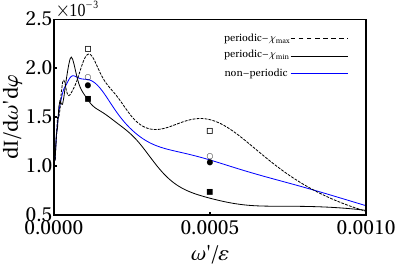}
  \caption{The emission spectra for the same parameters as in Fig.~\ref{fig:p_np_chi_spec}. The blue curve corresponds to the non-periodic case and the black solid and dashed lines correspond to the periodic case for specific values of $\varphi$ corresponding to the minimum $\chi_{min}$ and the maximum $\chi_{max}$, respectively. The circles (squares) designate the numerical calculation of a finite pulse of 10 (5) cycles. The filling (open) markers are related to $\varphi$ for $\chi_{min}$($\chi_{max}$).  }
  \label{fig:p_np}
  \end{center}
\end{figure}

Now we examine numerically the spectrum obtained for the non-periodic case, but with finite number of cycles in the laser pulse (as compared to the infinite pulse assumed by the analytical derivation). For the numerical calculations, we have evaluated Eq.~(\ref{eq:av BK}) numerically, employing the numerical trajectory for the electron, as for realistic laser pulses the trajectory is not available analytically. In Fig.~\ref{fig:p_np} full (hollow) circles designate 10 cycles with $\varphi$ corresponding to, respectively, the miminum (maximum) $\chi$ and full (hollow) squares are for 5 cycles with the same $\varphi$. First of all, the non-periodic spectrum, which represents averaging over $\varphi$, lies indeed in the middle between those curves, as expected. Secondly, one may see that the spectra for the finite laser pulse are far from the infinite pulse calculation. Moreover, the shorter the pulse is, the closer the results are to the periodic case. The reason is that the averaging out of the azimuthal dependence, as explained above,
requires many cycles of interaction.
The criterion which determines when one may employ the periodic formula is that the
 $\chi$-averaging is not significant, namely
\begin{equation}
 N (n-n_*) \ll 2 \pi
\label{eq:av dephase}
\end{equation}
where $N$ is the number of cycles in the laser pulse, $ n=\omega_2/\omega_1$ and $n_*$ is the closest integer number to $n$. For the parameters considered above this quantity reads 0.5 and 1, respectively. Consequently, the periodic expression provides a good estimation to the final result for short laser pulse, provided that the condition (\ref{eq:av dephase}) is fulfilled.

It is worth to point out that there is a certain regime where the emission for $\chi_{min}$ is larger than for $\chi_{max}$ in Fig.~\ref{fig:p_np_chi_spec}(d). This is because in the region of $\chi_{min}$ along the electron's trajectory, the emission of the electron is not uniquely determined by the quantum parameter $\chi$ as commonly believed. This is because of the violation of the local constant field approximation. In the formation length around $\chi_{min}$, $\chi(t)$ changes rapidly and increases up to the order of $\chi_{max}$ and thus the emission is also similar  near $\chi_{max}$ or even larger, see more discussions in Ref.\cite{lv2021_anomalous}.

\begin{figure}[b]
  \begin{center}
  \includegraphics[width=0.6\textwidth]{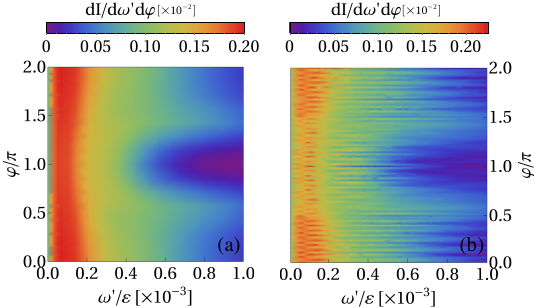}
  \caption{The emission spectra for the non-periodic case (a) and periodic case (b). Here we have $p_x=0.25m$, being about $0.25\%$ of the total energy, for both cases. The non-periodic case is for $\omega_2/\omega_1 \approx 60.2$, while the periodic case is for $\omega_2/\omega_1=60$. The other parameters are the same as in Fig.~\ref{fig:p_np_chi_spec}. }
  \label{fig:p_np_phi}
  \end{center}
\end{figure}

\textit{Non-periodic case versus periodic case with $p_x \neq 0$.}
Previously, we have discussed the emission of an electron in counterpropagating waves with vanishing transverse momentum. However, in a realistic experimental setup, the electrons in a beam always have non vanishing transverse momentum 
because of the angle spreading of the beam. In order to study the influence of the transverse momentum on the radiation process, we have in this section calculated the emission spectrum of an electron with $p_x \neq 0$ for both non-periodic and periodic cases.

In Fig.~\ref{fig:p_np_phi}, the spectra for $p_x$ being $0.25\%$ of the total energy have been investigated. Both of the spectra are not sysmetric with respect to the azimutal angle $\varphi$ as the $x-$direction is favorable. For the non-periodic case, even the gradual shift of $\chi$ regarding the azimuthal angle still happens for $p_x \neq 0$, the spectrum is nevertheless $\varphi$ dependent because the transverse momentum breaks the sysmmetry. Furthermore, the spectrum for the periodic case with nonzero $p_x$ has fringes with respect to $\varphi$. This means that the quantum parameter $\chi$ still has the similar dependence on the azimuthal angle like in Fig.~\ref{fig:p_np_chi_spec}(b).

\subsection{Validity condition}
\label{sec:validity}
In the following we 
derive the validity conditions for the emission formula obtained in the previous section. For this purpose, we recall that the next order correction to the trajectory employed in this paper reads
\begin{eqnarray}
\phi_1   \rightarrow   \phi_1 + C_1 \sin \phi_2 - C_{12} \sin (\phi_1-\phi_2) \quad
\phi_2   \rightarrow   \phi_2 + C_2 \sin \phi_1 + C_{12} \sin (\phi_1-\phi_2)
\end{eqnarray}
Let us substitute these modifications into the expression (\ref{eq:av psi}) for the phase $\psi$ and examine the additional terms.
Next, we take advantage of the identity
\begin{equation}
\sin \left(
\alpha \tau + \beta_1 \sin \kappa_1 \tau +
\beta_2 \sin \kappa_2 \tau - \varphi \right)
 = \sum_{s_1} \sum_{s_2} J_{s_1}(\beta_1) J_{s_2}(\beta_2)
 \sin \left[ \left( \alpha + s_1  \kappa_1 + s_2  \kappa_2 \right) \tau - \varphi \right]
\label{identity4}
\end{equation}
Please note that $\kappa$ here is just a parameter and not related to Eq.~(\ref{deriva}). Consequently, $\sin,\cos$ functions in (\ref{eq:av psi}) are replaced according to
\begin{eqnarray}
\cos \alpha \tau &\rightarrow& \mathcal{I}_1(\Lambda) \equiv \alpha \int{d \tau}  \sin \left(
\alpha \tau + \beta_1 \sin \kappa_1 \tau + \beta_2 \sin \kappa_2 \tau - \varphi\right)\,, \\
\sin \alpha \tau &\rightarrow& \mathcal{I}_2(\Lambda) \equiv \alpha \int{d \tau}  \cos \left(
\alpha \tau + \beta_1 \sin \kappa_1 \tau + \beta_2 \sin \kappa_2 \tau - \varphi \right)
\end{eqnarray}
with $\Lambda$ denoting $\left( \alpha,\beta_1,\kappa_1,\beta_2,\kappa_2,\varphi \right)$. Using (\ref{identity4}) one obtains
\begin{eqnarray}
\mathcal{I}_1 &=& -\sum_{s_1} \sum_{s_2} \frac{J_{s_1}(\beta_1) J_{s_2}(\beta_2) }{1+\left(s_1  \kappa_1 + s_2  \kappa_2 \right)/\alpha}
 \cos \left[ \left( \alpha + s_1  \kappa_1 + s_2  \kappa_2 \right) \tau - \varphi \right]\,, \\
\mathcal{I}_2 &=& \sum_{s_1} \sum_{s_2} \frac{J_{s_1}(\beta_1) J_{s_2}(\beta_2) }{1+ \left( s_1  \kappa_1 + s_2  \kappa_2 \right) / \alpha}
 \sin \left[ \left( \alpha + s_1  \kappa_1 + s_2  \kappa_2 \right) \tau - \varphi \right].
\end{eqnarray}
In the previous section the trigonometric identity $z_1^x \cos \phi_1+
z_1^y \sin \phi_1=
z_1 \sin (\phi_1-\varphi_1)
$ was employed, where $z_1,\varphi_1$ are given by (\ref{z1z2})
respectively. Analogously, in this case we have
\begin{equation}
z_1^x \mathcal{I}_2(\Lambda_0)+
z_1^y \mathcal{I}_1(\Lambda_0)=
z_1 \mathcal{I}_2(\Lambda_1)
\end{equation}
where
\begin{eqnarray}
\Lambda_0 = \left( \frac{\omega_1 \varepsilon}{m},C_1,\frac{\omega_2 \varepsilon}{m},-C_{12},\frac{(\omega_1-\omega_2) \varepsilon}{m}, 0 \right) \,, \quad 
\Lambda_1 = \left( \frac{\omega_1 \varepsilon}{m},C_1,\frac{\omega_2 \varepsilon}{m},-C_{12},\frac{(\omega_1-\omega_2) \varepsilon}{m}, \varphi_1 \right)
\label{Lambda1}
\end{eqnarray}
As a result, the modified phase may be written as
\begin{equation}
\psi= \psi_{np} \tau
-z_1 \mathcal{I}_2(\Lambda_1)
-z_2 \mathcal{I}_2(\Lambda_2)
-z_3 \mathcal{I}_2(\Lambda_3)
\label{eq:av psi_fm}
\end{equation}
where
\begin{eqnarray}
\Lambda_2 = \left( \frac{\omega_2 \varepsilon}{m},0,0,C_{12},\frac{(\omega_1-\omega_2) \varepsilon}{m}, \varphi_2 \right) \,, \quad
\Lambda_3 = \left( \frac{(\omega_1-\omega_2) \varepsilon}{m},C_1,\frac{\omega_2 \varepsilon}{m},-2C_{12},\frac{(\omega_1-\omega_2) \varepsilon}{m}, 0 \right) \,. \label{Lambda3}
\end{eqnarray}
Let us estimate the neglected  contribution to the phase, namely the difference between (\ref{eq:av psi_f}) and (\ref{eq:av psi_fm}). For the sake of simplicity, we split the corrections to 3 contributions, $\Delta \psi_1,\Delta \psi_2,\Delta \psi_3$ associated with $z_1,z_2,z_3$, respectively.
\begin{equation}
\Delta \psi = \Delta \psi_1 + \Delta \psi_2 + \Delta \psi_3 \,.
\end{equation}
In explicit terms, the corrections take the form $\Delta \psi_1= -z_1 \left[ \mathcal{I}_2\left(\Lambda_1 \right)-\sin \left( \phi_1 - \varphi_1 \right) \right]$,
and for $\Delta \psi_2,\Delta \psi_3$ we have $z_1 \rightarrow z_2,z_3$.
Since $C_1, C_{12} \ll 1$, we consider only first order corrections, namely $s_1=0,s_2=\pm 1$ and $s_1=\pm 1,s_2=0$. Therefore, one readily obtains
\begin{equation}
\mathcal{I}_2\left(\Lambda_1 \right)-\sin \left( \alpha \tau - \varphi \right) \approx
\frac{\beta_1}{2} \biggl( \frac{1}{1+\kappa_1/\alpha} \sin \left[ \left( \alpha + \kappa_1 \right) \tau - \varphi \right] - \frac{1}{1-\kappa_1/\alpha} \sin \left[ \left( \alpha - \kappa_1 \right) \tau - \varphi \right] \biggr)
 + \left( \beta_1,\kappa_1 \rightarrow \beta_2, \kappa_2 \right)
 \label{I_approx}
\end{equation}
where $J_1(\beta)=-J_{-1}(\beta) \approx \beta/2$ was used. Using Eqs.~(\ref{I_approx}), (\ref{Lambda1}) 
yields
\begin{equation}
\Delta \psi_1 = -\eta_1 \left[ \sin (\phi_1+\phi_2-\varphi_1) + \sin (\phi_1-\phi_2- \varphi_1) \right] + \eta_2 \left[ \sin (\phi_2-\varphi_1) - \sin (2 \phi_1-\phi_2 - \varphi_1) \right]
\end{equation}
where $\omega_1/\omega_2 \ll 1$ was employed. Analogously, for the other contribution one finds
\begin{eqnarray}
\Delta \psi_2 &=& -\eta_3  \sin ( \phi_1 -\varphi_2) - \eta_4 \sin (2 \phi_2-\phi_1-\varphi_2) \,, \\
\Delta \psi_3 &=& -\eta_5 \sin (\phi_1-2 \phi_2) + \eta_6 \sin \phi_1 + \eta_7 \sin \left[ 2 (\phi_1-\phi_2) \right] +\eta_8 \tau
\end{eqnarray}
The following coefficients were defined
\begin{eqnarray}
\eta_1 &\equiv& \frac{z_1 C_1 \omega_1}{2 \omega_2}\,,  
\eta_2 \equiv \frac{z_1 C_{12} \omega_1}{2 \omega_2}\,,  
\eta_3 \equiv \frac{z_2 C_{12} \omega_2}{2 \omega_1}\,,  
\eta_4 \equiv \frac{z_2 C_{12} }{4}\,,  \\
\eta_5  &\equiv&  \frac{z_3 C_1 }{4}\,,  
\eta_6 \equiv \frac{z_3 C_1 \omega_2}{2 \omega_1}\,,  
\eta_7 \equiv \frac{z_3 C_{12} }{2 }\,, 
\eta_8 \equiv \frac{(\omega_1-\omega_2) \varepsilon z_3 C_{12}}{m}\nonumber  \,. 
\end{eqnarray}
In order to formulate the general validity condition, we notice that the phase (\ref{eq:av psi_f}) contains a linear term with low ($\omega_1$) and high ($\omega_2, \omega_2-\omega_1$) frequencies. Therefore, we require that the coefficients of the high frequency corrections, will be smaller as compared to $z_2, z_3$. Similarly, the coefficients of the low frequency should be lower than $z_1$, and the one corresponding to the linear term smaller than $\psi_{np}$. Hence the general validity condition may be cast in the form
\begin{eqnarray}
\eta_1,\eta_2,\eta_4,\eta_5,\eta_7   \ll   z_2,z_3 \quad
\eta_3,\eta_6   \ll   z_1 \label{eq:av validity_general} \quad
\eta_8   \ll   \psi_{np}  \,.
\end{eqnarray}
We call attention to the fact that these conditions depend on the emitted photon properties $\omega, \theta, \varphi$. As a result, for given interaction parameters (laser amplitudes, particle energy), part of the spectrum may be described by our analytical expression whereas a different part may exhibit deviations. Hence, one should verify that (\ref{eq:av validity_general}) holds for the entire spectral range of interest. In the strong field case, however, the situation is much simplified and simple criteria are derived, which hold for the entire spectrum.

Let us consider explicitly the strong field regime ($\xi_1 \gg 1$).
As explained in Sec.~\ref{sec:strongField}, in this regime the emission is restricted to a limited angle range, for which the Bessel coefficients may be approximated by $z^c_1,z^c_2,z^c_3$. Substituting these expressions to the requirement (\ref{eq:av validity_general}) and employing the trajectory validity conditions in Sec.~\ref{sec:erranaly} as well as the approximation $\frac{\omega_2}{\omega_1} = \frac{1+\bar{v}_z}{1-\bar{v}_z} \approx \frac{4 \varepsilon^2}{m_*^2}$, we find that $\eta_4,\eta_5,\eta_7, \eta_8$ obey (\ref{eq:av validity_general}) by definition.
Employing Eq.~\eqref{eq:av Zc} as well as the expressions for $C_1,C_{12}$, the validity condition is simplified to
\begin{eqnarray}
\frac{\eta_1}{z_2} = \frac{p_x \xi_1}{2 m_* \xi_2} \ll 1, \quad \frac{\eta_6}{z_1}=\frac{p_x \xi_2}{2 m_* \xi_1} \ll 1 \, , \quad
\frac{\eta_2}{z_2} = \frac{m^2 \xi_1^2}{2 \varepsilon} \ll 1, \quad \frac{\eta_3}{z_1}=\frac{m^2 \xi_2^2}{2 \varepsilon} \ll 1\, . \label{eq:av reduced_cond2}
\end{eqnarray}
The last three conditions are automatically fulfilled according to the validity conditions for the trajectory. Hence, only a single additional condition, corresponding to $\eta_1$ in Eq.~(\ref{eq:av reduced_cond2}), is required to validate the applied formalism:
\begin{equation}
\frac{p_x }{2 m \xi_2} \ll 1.
\end{equation}

\begin{figure}[b]
  \begin{center}
  \includegraphics[width=0.35\textwidth]{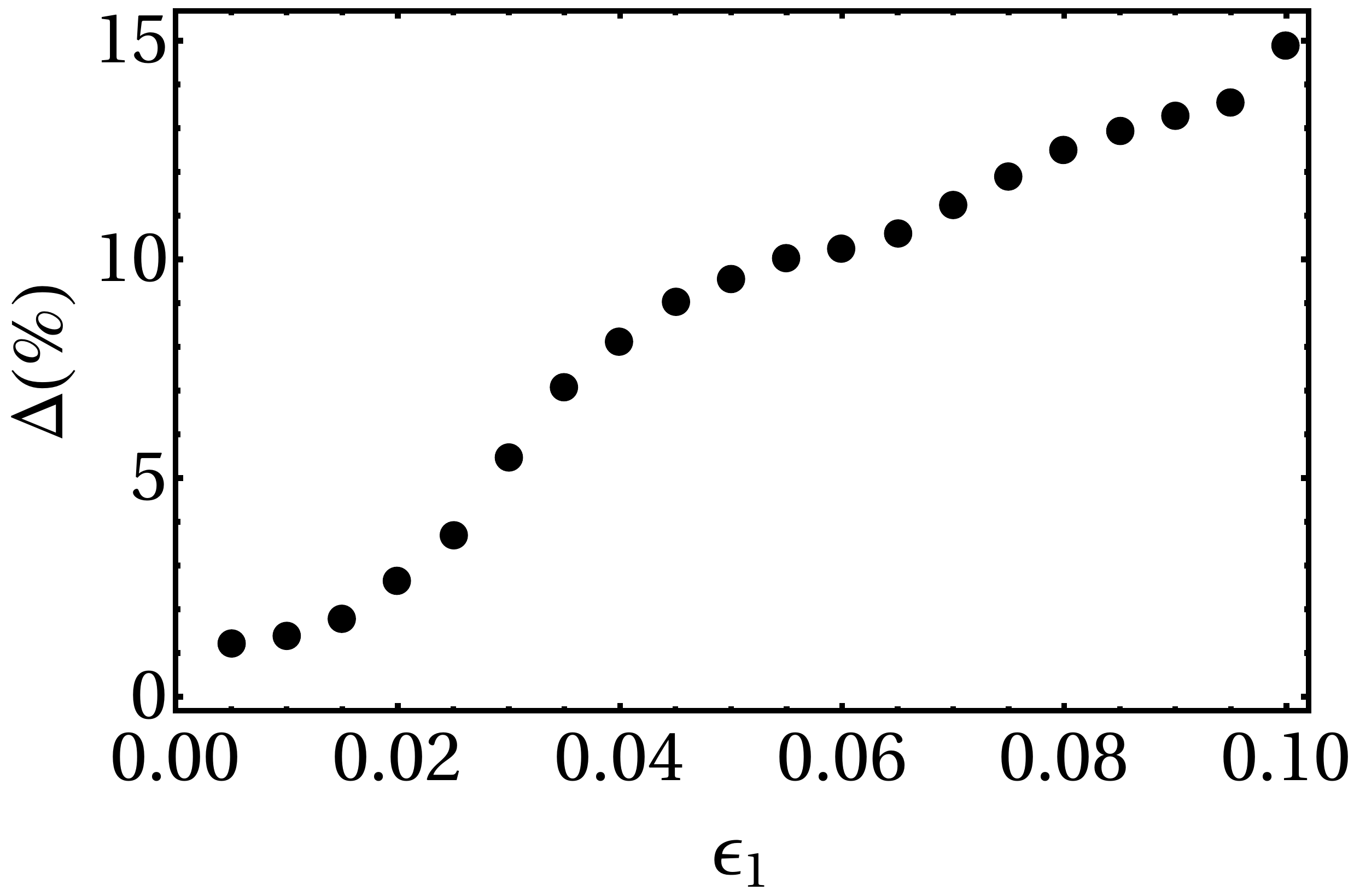}
  \caption{ The relative deviation in percentage between the analytical and the numerical spectrum as a function of the small parameter $\epsilon_1$. See text for details.}
  \label{fig:validity}
  \end{center}
\end{figure}
As demonstrated above, the analytical approximation depends on several criteria being fulfilled.
In the following, we examine in detail the strong field case, where the number of quantities required to be low is relatively small, allowing for a tractable study of the error. The main quantities, which stem from the trajectory approximation, are $\epsilon_1,\epsilon_2$ given in Sec.~\ref{sec:erranaly}.  In the following we investigate systematically and quantitatively the relation between these parameters and the corresponding error. For the sake of this purpose, a new quantity is introduced
\begin{equation}
\Delta=\frac{\Delta_1+\Delta_2+\Delta_3}{3},
\quad
\Delta_i \equiv 100 \frac{|I_n(u_i)-I_a(u_i)|}{I_n(u_i)}
\end{equation}
measuring the relative difference $\Delta $ (in percent) between the analytical ($I_a$) and numerical ($I_n$) results. It is evaluated in 3 points $u_i$ on the spectrum.  Fig.~\ref{fig:validity} shows the deviation $\Delta$ as a function of $\epsilon_1$, assuming that $\epsilon_2$ vanishes.
One can see that error grows monotonically, and that $\epsilon_1=0.1$ yields a deviation of $15\%$.
In order to examine the influence of $\epsilon_2$, the calculation was generalized to two dimensions, and the results are presented in Tab.~\ref{Tab:table2}. One may see that the impact of $\epsilon_2$ is significantly smaller as compared to $\epsilon_1$. The results presented in this subsection provides quantitative information which may be valuable when applying our expressions in practice.
\begin{table}[H]
  \centering
  \begin{tabular}{|c|c|c|c|c|c|}
      \hline
      \diagbox{$\epsilon_1$}{$\epsilon_2$} & 0.03 & 0.06 &0.09& 0.12& 0.15 \\
      \hline
      0.03 & 3.01 & 10.25 & 13.85 & 18.35 & 25.46 \\
      \hline
      0.06 & 3.05 & 10.60 & 14.19 & 19.10 & 26.19 \\
      \hline
      0.09 & 3.10 & 11.00 & 14.72 & 19.83 & 26.99 \\
      \hline
      0.12 & 3.19 & 11.40 & 15.57 & 20.56 & 27.81 \\
      \hline
      0.15 & 3.32 & 11.80 & 16.79 & 21.45 & 28.90 \\
      \hline
  \end{tabular}
  \caption{The relative difference between the analytical and numerical results as a function of
the parameters $\epsilon_1$ and $\epsilon_2$.}
  \label{Tab:table2}
\end{table}

\subsection{Realistic pulse effects}
\label{sec:realistic}
The analytical derivation presented above assumes that the laser fields are monochromatic plain waves. This approximation is appropriate for long pulses (dozens of cycles) which are focused on large spots (radius of dozens of wavelengths). However, realistic pulses tend to be short and tightly focused, in order to maximize the obtained intensity for a given pulse energy. Therefore, for practical reasons it is highly important to thoroughly examine the dependence of the emission on pulse duration and focal size. In particular, we wish to establish qualitatively which spectral features are affected by shortening / focusing the laser pulse, what is the amplitude of the deviation and to find the conditions for which the spectrum recovers the analytical result.

In order to specify the spatial and temporal shape for the realistic laser pulse, the following quantities are introduced
\begin{equation}
  X \equiv \frac{x}{w_0} \,, 
  Y \equiv \frac{y}{w_0} \,, 
  Z \equiv \frac{z}{z_r} \,,  
  z_r \equiv \frac{\omega w_0^2}{2} \,,
  f \equiv \frac{i}{i+Z} \,,  
  \rho \equiv \sqrt{X^2+Y^2}\,.
\end{equation}
The vector potential corresponding to this pulse reads \cite{Salamin_2002}
\begin{equation}
  A_x = A_0 g (k \cdot x) f e^{-f \rho^2} e^{i k \cdot x}  
  \left( 1+ \epsilon^2 \left[ \frac{f}{2}- \frac{f^3 \rho^4}{4} \right]
  +\epsilon^4 \left[ 
    \frac{3 f^2}{8} - \frac{3 f^4 \rho^4}{16}- \frac{f^5 \rho^6}{8}+ \frac{f^6 \rho^8}{32}
    \right]
  \right)   
\label{eq:av A_sp}
\end{equation}
where $A_0$ denotes the amplitude.  Since we consider a circular polarization, the $y$ component is given by $A_y=i A_x$.  The electromagnetic fields can thus be derived from the above vector potential by
\begin{equation}
  \label{eq:EandB}
    \vec{E}= -i \omega \vec{A} - \frac{i}{\omega} \pmb{\nabla}
               \left( \pmb{\nabla} \cdot \vec{A} \right) \,, 
    \vec{B}= \pmb{\nabla} \times \vec{A} \,.
\end{equation}

\begin{figure} 
  \begin{center}
  \includegraphics[width=0.45\textwidth]{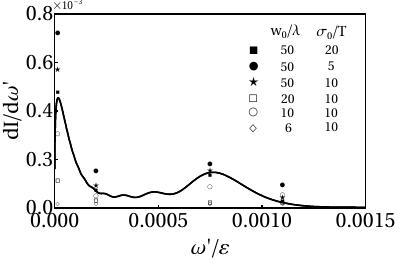}
  \caption{The emitted spectra for various pulse duration and spot sizes. $w_0$ is the spot radius and $\sigma_0$ the pulse duration. Here $w_0$ and $\sigma_0$ are presented in the unit of laser wave length $\lambda$ and period $T$, respectively. The solid line stands for the analytical expression ($w_0,\sigma_0 \rightarrow \infty$).}
  \label{fig:reallistic}
  \end{center}
\end{figure}

In the calculation below, we choose a $\sin^2$-function with $\sigma_0$ denoting the pulse length for $g(k \cdot x)$ as the temporal envelope. Fig.~\ref{fig:reallistic} depicts the angle integrated emission of a particle interacting with pulses with normalized amplitudes $\xi_1=12.5, \xi_2=0.1, \varepsilon=80 m$, respectively. The solid line stands for the analytical expression. Numerical calculations corresponding to variety of pulse durations  and focal radii were carried out as well. For the sake of comparison, we wanted to keep the energy of the particle in the main part of the pulse identical for all compared cases. For this purpose, the initial electron energy was zero and its initial location $z_0$ was tuned, namely the distance to the beginning of the $\xi_1$ and $\xi_2$ beams. From an experimental point of view, it may be realized by placing atoms which are ionized by the laser field.

As expected, the analytical formula coincides fairly well with the numerical calculation for a long pulse with large focus $w_0=50\lambda, \sigma_0=20T$, with $\lambda$ and $T$ as the laser wave length and laser period, respectively. Let us examine the influence of the temporal width first. Decreasing the duration to $\sigma=10T$ does not change much the spectrum. However, for ultrashort pulses (full circles) with $\sigma=5T$, the emission significantly increases. This may be explained by the fact that the rapid rise of the pulse is accompanied by stronger acceleration and enhanced $\chi$ value. Moreover, the shorter the pulse is, the larger the edge effect will be in the emission spectrum. This edge effect will induce deviations of the spectrum from the LCFA predictions and enhance the emission, especially in the high energy domain \cite{lv2021_anomalous}.

As for the spatial focusing, one observes an opposite trend. Namely, a small spot results in a significant decrease in the emitted spectrum, as well as in a deformation of its spectral shape. We suggest that this outcome stems from the fact that tightly focused beams rapidly expel the particle from the focus due to the transverse pondermotive force. Furthermore, one may notice that even moderate focusing, $w_0=20\lambda$, results in a considerable deviation from the one dimensional case. Thus, Fig.~\ref{fig:reallistic} shows that finite duration yields significant deviation from the analytical expression only for ultrashort pulses, whereas the focal radius has greater influence and should be fairly large in order to recover the theoretical result.

\section{Summary and conclusion}
\label{sec:con}
We have investigated the radiation properties of a relativistic electron in counterpropagating laser waves within the semiclassical formalism introduced by Baier and Katkov. This formalism is valid when the electron dynamics in the background classical fields is  quasiclassical. It treats a photon emission  quantum mechanically, fully taking into account the quantum recoil of the emitted photon. As the formalism employs the electron classical trajectory in the given fields, we firstly investigate in detail the electron classical dynamics in the  counterpropagating laser beam setup.  
The classical momentum and trajectory are analytically derived assuming that the particle energy is the dominant scale and that the angle between the particle propagation direction and the beams axis is small (see the exact conditions 
in Eq.~(\ref{eq:av cirterion3F})). The trajectory characteristics as a function of the laser parameters and the particle energy are discussed. In particular, we show that in the case when the  quantum parameters induced by each of the beams are comparable, $\chi_1 \approx \chi_2$, a peculiar spike-like feature arises. Since its typical time scale is significantly shorter as compared to $1/\omega_1,1/\omega_2$, it will bear great significance to the corresponding radiation properties. Moreover, a detailed comparison with the full numerical solution was carried out resulting in a good agreement and validating our analytical solution in the given conditions. The dependence of the small deviations with respect to the exact solution on the  parameters has been systematically investigated. We have observed an interesting relationship of the cycle-averaged momentum in the field to the asymptotic one. We show that the final average momentum depends on the order by which the laser beams are turned on.

Further, employing the approximated analytical trajectory, the radiation has been calculated in the Baier-Katkov semiclassical framework.
The Baier-Katkov integrals were analytically solved yielding closed formulas in terms of sums over Bessel functions. Different regimes, periodic and non-periodic, are explored. 

We concentrated on the strong field regime, which was found to be of particular interest for anomalous LCFA violation
\cite{lv2021_anomalous}. An optimised calculation method based on a physical reasoning is suggested, which enables quick summation over the numerous Bessel harmonics appearing in the analytical formula. The result is employed to compare in detail the periodic and the non-periodic regimes. We have observed that as opposed to the non-periodic case, where non-uniformity in the azimuthal direction finally averages to zero, in the periodic case considerable dependence on the azimuthal angle appears. We found that in a rather short laser pulse the emission in the non-periodic case becomes similar to the periodic one.
 Furthermore, we 
 analyze numerically the effect introduced by a finite duration and spot size of the beams, which are not included in the analytical derivation. We demonstrate that the ultrashort pulse results in enhanced emission while tightly focused beam reduce the emitted energy and give physical explanations.

Finally, elaborated analytical analysis of the validity condition is presented. In the general case, it depends on the energy and angle of the emitted photon. In the strong field case, it reduces to a simple restriction on the ratio between the energy and the laser amplitude. The error in the spectrum is evaluated numerically and systematically explored as a function of the small quantities lying in the foundation of the theoretical approximation.

\section*{Acknowledgment}

Q.Z.L and E.R. contributed equally to the work, to numerical and analytical calculations, respectively.
E.R. acknowledges partial support from the Alexander von Humboldt Foundation.

\bibliography{lcfa}

\end{document}